\documentclass[12pt,a4paper]{article}
\usepackage{amsmath,amssymb,extarrows,mathtools,graphicx,subfigure,setspace}
\usepackage[all]{xy}
\usepackage{epsfig,comment,geometry,comment,color,cite,slashed}
\usepackage[active]{srcltx}
\usepackage[affil-it]{authblk}
\usepackage{fullpage}
\numberwithin{equation}{section}
\newcommand{\be}{\begin{equation}}
\newcommand{\ee}{\end{equation}}
\newcommand{\bea}{\begin{eqnarray}}
\newcommand{\eea}{\end{eqnarray}}
\newcommand{\nn}{\nonumber}

\newcommand{\LO}{{(0)}}
\newcommand{\FO}{{(1)}}
\newcommand{\SO}{{(2)}}
\newcommand{\TO}{{(3)}}

\newcommand{\si}{\sigma}
\newcommand{\tb}{\textbf}

\begin{document}
\begin{titlepage}
{\title{\bf\fontsize{14}{15.2}{Holographic Two-Point Functions in Conformal Gravity}}}
\vspace{.5cm}
\author[a]{Ahmad Ghodsi \thanks{a-ghodsi@ferdowsi.um.ac.ir}}
\author[b]{Behnoush Khavari \thanks{b.k@ipm.ir}}
\author[c]{Ali Naseh \thanks{naseh@ipm.ir}}
\vspace{.5cm}
\affil[a]{ Department of Physics, Ferdowsi University of Mashhad,    
\hspace{5.5cm} P.O.Box 1436, Mashhad, Iran}
\affil[b]{ School of Physics, Institute for Research in Fundamental
Sciences (IPM), 
\hspace{5.5cm} P.O.Box 19395-5531, Tehran, Iran}
\affil[c]{ School of Particles and Accelerators, Institute for Research in Fundamental
Sciences (IPM), 
\hspace{.001cm} P.O.Box 19395-5531, Tehran, Iran}
\renewcommand\Authands{ and }
\maketitle
\vspace{-12cm}
\begin{flushright}
{\small
{\tt arXiv:yymm.nnnn} \\
%IPM/P-2014/xxx \\
}
\end{flushright}
\vspace{10cm}
%%%%%%%%%%%%%%%%%%%%%%%%%%%%%%%%%%%%%%%%%%%%%%%%%%%%%%%%%%%
\begin{abstract}
In this paper we compute the holographic two-point functions of four dimensional conformal gravity. Precisely we calculate the two-point functions for Energy-Momentum (EM) and Partially Massless Response (PMR) operators that have been identified as two response functions for two independent sources in the dual CFT. 
The correlation function of EM with PMR tensors turns out to be zero which is expected according to the conformal symmetry. The two-point function of EM is that of a transverse and traceless tensor, and the two-point function of PMR which is a traceless operator contains two distinct parts, one for a transverse-traceless tensor operator and another one for a vector field, both of which fulfill criteria of a CFT. We also discuss about the unitarity of the theory.
\end{abstract}
\end{titlepage}
%%%%%%%%%%%%%%%%%%%%%%%%%%%%%%%%%%%%%%%%%%%%%%%%%%%%%%%%%%%
\setcounter{footnote}{0}
\addtocontents{toc}{\protect\setcounter{tocdepth}{4}}
\setcounter{secnumdepth}{4}
%\tableofcontents
%%%%%%%%%%%%%%%%%%%%%%%%%%%%%%%%%%%%%%%%%%%%%%%%%%%%%%%%%%%%
\section{Introduction}
	Since its formulation, the AdS/CFT correspondence \cite{Maldacena:1997re,Witten:1998qj,Gubser:1998bc}  has been used as a strong tool for non-perturbative calculations, among which the well-studied subject of holographic renormalization 
is of great importance \cite{Skenderis,Bianchi:2001kw,deBoer,deHaro,Papadimitriou:2004ap,Papadimitriou:2005ii,Karch:2005ms,Skenderis:2008dh,Kanitscheider:2008kd}. One of the aims of holographic renormalization is to find the holographic $n$-point functions, which according to the holographic prescription, are assumed to be renormalized $n$-point functions of operators of a Quantum Field Theory (QFT) which is dual to the gravitational theory under study. 

Consider a $d+1$-dimensional theory of gravity which has an $AdS_{d+1}$ vacuum solution. The fluctuations around this vacuum may have some nonvanishing values on the boundary of the $AdS_{d+1}$ space-time. According to the $AdS_{d+1}/CFT_{d}$ dictionary these can be interpreted as sources for the QFT operators living on the $d$-dimensional boundary space\cite{Gubser:1998bc,Witten:1998qj}.

One can find the correlation functions of these operators if the generating functional of the QFT is known. This information is provided by the partition function of the gravitational theory which has been argued to be the same as the QFT generating function \cite{Witten:1998qj}. In the approximation of classical supergravity, to the first order in the saddle point approximation, the gravity partition function is given by its on-shell action, which is a functional of the boundary values of the fields. In this regime,  on-shell supergravity action is equal to the generating function of QFT connected graphs. 

Having the generating functional in hand, one can calculate the one-point function in the presence of the source, and then by further varying the one-point function with respect to the source, it is possible to obtain higher $n$-point functions.  Actually, $n$-point functions in a QFT suffer from  UV divergences and therefore require a renormalization process. To do this, one needs to regularize the on-shell action and add some counter-terms in a covariant way, so that the on-shell action becomes finite and the $n$-point functions which are extracted from it, are renormalized.  For more detailed discussions see for example \cite{Skenderis}.

This analysis has been done for different types of gravitational theories. For cosmological Einstein-Hilbert gravity this is well done in \cite{deHaro,Balasubramanian:1999re} and for some higher derivative theories in \cite{Skenderis:2009nt,Hohm:2010jc,Alishahiha:2010bw, Afshar:2011qw, Johansson:2012fs}, where the one-point and two-point functions of dual operators have been calculated. 

One of the interesting higher derivative gravitational theories which has been recently studied by using the holographic renormalization is the conformal gravity (CG)\cite{Grumiller:2013mxa}. Conformal gravity action is given by a Weyl squared term as 
	\be\label{eqCG2}
	I_{CG} = \alpha_{CG} \int d^4 x \sqrt{|g|} C^{\alpha\beta\gamma\delta} C_{\alpha\beta\gamma\delta}\,,
	\ee
which is invariant under the Weyl transformation $g_{\mu \nu}\rightarrow e^{2\Omega(x)}g_{\mu \nu}$ and $\alpha_{CG}$ is a dimensionless coupling constant. A positive definite Euclidean quantum action and an acceptable Newtonian limit is possible only  for $\alpha_{CG} > 0$\cite{Hasslacher:1980hd}\footnote{In this letter we assume that $\alpha_{CG}=1$.}. The equation of motion extracted from this action is known as the Bach's equation \cite{Bach}. It is well known that the solutions of Einstein gravity, and as a specific example the $AdS$ space-time, are also solutions of the Bach's equation. Moreover, because of the higher derivative nature of CG, the Bach's equation admits solutions which are not the Einstein spaces. 

In contrast with the Einstein gravity which is not renormalizable \cite{Goroff:1985sz, Goroff:1985th}, CG is power-counting renormalizable \cite{Stelle:1976gc} (in fact, asymptotically free\cite{Hasslacher:1980hd, Julve:1978xn,Fradkin:1981iu}) and therefore is considered as a possible UV completion of gravity \cite{Adler:1982ri}. Generally CG contains the ghost fields because of its higher derivative terms \cite{Tomboulis:1983sw, Boulware:1983td}. In recent years, it has been in the center of attention. It is used to explain the galactic rotation curves without need for dark matter \cite{Mannheim:1988dj,Mannheim:2005bfa,Mannheim:2010ti,Mannheim:2011ds}. Moreover it is emerged in the context of the Gauge/Gravity duality as a counter term \cite{Henningson:1998gx,Liu:1998bu} and also arises in the twistor string theory \cite{Witten:2003nn, Berkovits:2004jj}. It has been shown that it is equal (at the linearized level) to the Einstein gravity by imposing a special boundary condition\cite{Maldacena:2011mk}. The appearance of the CG in quantum gravity context has been studied in \cite{tHooft:2011aa}.

More recently, it has been shown that the on-shell action for the four dimensional conformal gravity is renormalized \cite{Grumiller:2013mxa}. Doing the near boundary analysis, it has been argued that the first two coefficients in the Fefferman-Graham (FG) expansion of the boundary metric can consistently be interpreted as two independent sources for two operators in the boundary theory. These operators are the Energy-Momentum (EM) tensor and the Partial Massless Response (PMR) for which the one-point functions have been worked out in \cite{Grumiller:2013mxa}. We will review this analysis in section 2. To push forward this calculation, in this paper we are interested in finding the two-point functions of the dual operators using the AdS/CFT conjecture.

In fact as far as one is concerned with the derivation of the renormalized action and one-point functions, the study can be limited to the near boundary survey; however, for higher $n$-point functions, one needs to know about the behavior of the solution inside the bulk and so, the asymptotic analysis is not sufficient any more \cite{deHaro}. That is what we are going to do in this paper in section 3. Nevertheless, we note that since in this study, we are just considering two-point functions, it is fortunately sufficient to deal with the linearized equations of motion for metric fluctuations. 
	
	That is the way we are going to perform our computation. We first linearize the Bach's equation in section 2 and then find its solutions in section 3. Equipped with these solutions, in addition to the one-point functions given in \cite{Grumiller:2013mxa}, we will finally calculate the desired two-point functions and verify that these two-point functions are consistent with the Ward identities obtained in \cite{Grumiller:2013mxa}. The last section is devoted to the conclusion. 
 %%%%%%%%%%%%%%%%%%%%%%%%%%%%%%%%%%%%%%%%%%%%%%%%%%%%%%%%%%%%
\section{Linear analysis of conformal gravity}
As has been argued in \cite{Grumiller:2013mxa}, the Fefferman-Graham expansion of the boundary metric in conformal gravity shows two independent sources for two operators in the boundary theory. In this section we will obtain this result by performing a simple analysis of the linearized equations of motion. Let us start from the action of four dimensional conformal gravity in (\ref{eqCG2}) and write it in terms of the curvature tensors  
\bea
S=\int d^{4}x \sqrt{g} \Big(\mathcal{R}_{\mu\nu\lambda\rho} \mathcal{R}^{\mu\nu\lambda\rho} -2 \mathcal{R}_{\mu\nu} \mathcal{R}^{\mu\nu} + \frac13 \mathcal{R}^2\Big)\,.
\eea
The equation of motion corresponding to this action (Bach's equation) is given by 
\be \label{EOM}
4 \mathcal{R}^{\alpha\beta} \mathcal{R}_{\mu\alpha\nu\beta} -\frac43 \mathcal{R} \mathcal{R}_{\mu\nu} +2\Box \mathcal{R}_{\mu\nu} -\frac23 \nabla_{\mu}{\nabla_{\nu}{\mathcal{R}}}
+ \big( \frac13 \mathcal{R}^2 - \mathcal{R}^{\alpha\beta} \mathcal{R}_{\alpha\beta} - \frac13 \Box \mathcal{R}\big) g_{\mu\nu} =0\,.
\ee
In the subsequent section we present a linearization analysis of the above equation of motion around $AdS_{4}$ background in the Fefferman-Graham (FG) coordinates {\cite{Fefferman} and then perform the asymptotic analysis. We have checked all results by using the Cadabra software \cite{Peeters:2006kp,Peeters:2007wn}.
%%%%%%%%%%%%%%%%%%%%%%%%%%%%%%%%%%%%%%%%%%%%%%%%%%%%%%%%%%%%%%%%%%%%%%%%%
\subsection{CG equations of motion in FG coordinates}\label{l.F.G.e.o.m}
To perform the linearized analysis, let us start from the Bach's equation (\ref{EOM}) in the FG coordinates. In this coordinate, the metric can be written as 
\begin{equation}
ds^2=\frac{dr^2}{4r^2} +\frac{1}{ r} h_{i j}(x, r) dx^i dx^j\,,
\end{equation}
where $h_{ij}$ in this notation contains a flat metric part plus the metric fluctuations.
By substituting this metric into the Bach's equation and up to the non-linear terms in metric fluctuations we find the following expressions for $ r r$, $ij$ and $i r$ components respectively\footnote{For more details such as the Christoffel symbols and curvature tensors in the FG coordinates see the appendix A.}
\begin{eqnarray}\label{eomrr}
r \big( R''\!-\! 2 R''_{a b} h^{a b} \!-\! \frac{2}{3} \Box{h''_{ab}} h^{ab}\big) \!-\! \frac{1}{12} \Box{R} \!-\! \frac{5}{12} R'_{a b} h^{a b} + 2 \nabla^{a}{\nabla^{b}{h'_{a b}}}
\!-\! \frac{7}{3} \Box{h'_{ab}} h^{ab} \!+\! \cdots =0\,,  
\eea
\bea
\label{eomij}
&&  r \big(   12 R'_{i j} \!-\! \frac{2}{3} \nabla_{i}{\nabla_{j}{R}} + 2 \Box{R_{i j}}\!-\! \frac{2}{3} R' h_{i j} \!-\! 4 \nabla_{i}{\nabla^{a}{h'_{j a}}} \!-\! 4 \nabla_{j}{\nabla^{a}{h'_{i a}}}\!+\! \frac{16}{3} \nabla_{i}{\nabla_{j}{h'_{a b}}} h^{a b} \!+\! 2 \Box{h'_{i j}}   \nn \\
&&- 12 h''_{i j}+ (  4 h^{a b} h''_{a b}  +\frac{2}{3} h^{ab} \Box{h'_{ab}} - \frac{1}{3} \Box{R}  ) h_{i j} \big) + r^{2} \big( 8 R''_{i j} - \frac{4}{3} R'' h_{i j} - 48 h'''_{i j} - 4 \Box{h''_{i j}}\nn \\
&&
+ h^{a b} (\frac{8}{3} \nabla_{i}{\nabla_{j}{h''_{a b}}} + 16  h'''_{a b} h_{i j}+\frac{4}{3} \Box{h''_{ab}} h_{i j})\big) +  r^{3} \big( \frac{16}{3} h^{a b} h''''_{a b} h_{i j} - 16 h''''_{i j} \big) +\cdots  =0\,, 
\eea
\bea
\label{eomir}
&&  r \big(\frac{1}{2} \Box{\nabla^{b}{h'_{i b}}}\! -\! \frac{1}{2} \Box{\nabla_{i}{h'_{b c}}} h^{b c} + 5 (\nabla^{a}{h'_{i a}})' \!-\! 5 (\nabla_{i}{h'_{a b}})' h^{a b} + 4 \nabla_{i}{h''_{a b}} h^{a b} \!-\! 2 \nabla^{a}{h''_{i a}} \! -\! \frac{1}{3} \nabla_{i}{R'} \big) \nn \\
&&+ r^{2} \big( 2 (\nabla^{a}{h'_{i a}})'' - 2 (\nabla_{i}{h'_{a b}})'' h^{a b} + \frac{4}{3} h^{a b} \nabla_{i}{h'''_{a b}} \big)+ \cdots =0\,,
\end{eqnarray}
where prime denotes the derivative with respect to the $ r$ coordinate. $\nabla$ and $\Box$ and all curvature tensors are defined in the boundary space, i.e.
 are constructed out of the boundary metric $h_{i j}$. Dots in the above equations represent the irrelevant terms (non-linear in boundary metric fluctuations).
%%%%%%%%%%%%%%%%%%%%%%%%%%%%%%%%%%%%%%%%%%%%%%%%%%%%%%%%%%%%%%%%%%%%%%%%%%%%%%%%%

To find the linearized equations of motion for metric fluctuations around the $AdS_{4}$ space-time we substitute $h_{i j}=\eta_{i j}+f_{i j}$ into the linearized equations (\ref{eomrr}), (\ref{eomij}) and (\ref{eomir}). In this way the $ r r$ component (\ref{eomrr}) is given by
\begin{equation}\label{l3}
 r\big(\frac{1}{3}\Box f''-\partial^{i}\partial^{j}f''_{i j}\big)-\frac{1}{2}\partial^{a}\partial^{b}f'_{a b}+\frac{1}{6}\Box f'+\frac{1}{12}\Box^2 f-\frac{1}{12}\partial^{a}\partial^{b}\Box f_{a b}=0\,,
\end{equation}
and the $ij$ component (\ref{eomij}) will be 	
\begin{eqnarray}\label{l1}
&& r^{3}\big(\frac{16}{3}f'''' \eta_{i j}-16f''''_{i j}\big)\!+\! r^{2}\big((16f''' +\frac{8}{3}\Box f'' -\frac{4}{3}\partial^{a}\partial^{b}f''_{a b}) \eta_{i j}
-48f'''_{i j}-8\Box f''_{i j} -\frac{4}{3}\partial_{i}\partial_{j}f''\nn \\
&&+4\partial_{i}\partial^{a}f''_{a j}+4\partial_{j}\partial^{a}f''_{a i}\big)\! +\!  r\big((4f''+\frac{4}{3}\Box f'-\frac{2}{3}\partial^{a}\partial^{b}f'_{a b}-\frac{1}{3}\partial^{a}\partial^{b}\Box f_{a b}+\frac{1}{3}\Box^2 f) \eta_{i j}\!-\!12f''_{i j}\nn \\
&&
-4\Box f'_{i j}+2\partial_{i}\partial^{a}f'_{a j}+2\partial_{j}\partial^{a}f'_{a i}-\frac{2}{3}\partial_{i}\partial_{j}f' +\partial_{i}\partial^{a}\Box f_{a j}+\partial_{j}\partial^{a}\Box f_{a i}-\frac{1}{3}\partial_{i}\partial_{j}\Box f -\Box^2 f_{i j}\nn \\
&&-\frac{2}{3}\partial_{i}\partial_{j}\partial^{a}\partial^{b}f_{a b}\big)=0\,.
\end{eqnarray}
Finally the $i r$ component (\ref{eomir}) simplifies to
\begin{equation}\label{l2}
2 r^{2}\big(\partial^{j}f'''_{i j}-\frac{1}{3}\partial_{i}f'''\big)+ r\big(3\partial^{j}f''_{i j}-\partial_{i}f''+\frac{1}{2}\partial^{a}\Box f'_{i a}-\frac{1}{6}\partial_{i}\Box f'-\frac{1}{3}\partial_{i}\partial^{a}\partial^{b}f'_{a b}\big)=0\,,
\end{equation}
where in these new equations $\Box=\partial_{k}\partial^{k}$.

Now to study the theory near the boundary, one can substitute in the above linearized equations, the following FG expansion for metric fluctuations $f_{i j}$, as has been presented in \cite{Grumiller:2013mxa}
	\begin{eqnarray}\label{fexpand}
	f_{i j}=f^{(0)}_{i j}+ r^{\frac{1}{2}}f^{(1)}_{i j}+  r f^{(2)}_{i j}+  r^{\frac{3}{2}}f^{(3)}_{i j}+\cdots\,,
	\end{eqnarray}
	where all the expansion coefficients $f^{(n)}_{i j}$ are allowed to depend on the boundary coordinates $x_{i}$. Actually, the validity of this FG expansion for the boundary metric can be directly verified from the exact solutions of the linearized Bach's equation that we are going to figure out in the subsequent sections. As we will see, solutions always have some series expansion in $\sqrt{ r}$ with integer powers. This clearly explains why we choose the FG expansion of (\ref{fexpand}) the way it is.
	
By this expansion the $ r r$ component is obtained as
\begin{eqnarray}\label{f04rr}
&& - \frac{1}{2} \partial^{a}{\partial^{b}{f^{(2)}_{a b}}}
+\frac{1}{6} \Box{f^{(2)}} -\frac{1}{12} {\partial^{a}{\partial^{b}{\Box f^{(0)}_{a b}}}} + \frac{1}{12} \Box^2{{f^{(0)}}} \nn \\
&& + r^{\frac{1}{2}} \big(-\frac{3}{2} \partial^{a}{\partial^{b}{f^{(3)}_{a b}}} +\frac{1}{2} \Box{f^{(3)}} - \frac{1}{12} {\partial^{a}{\partial^{b}{\Box f^{(1)}_{a b}}}}
+\frac{1}{12} \Box^2{{f^{(1)}}}\big) \nn \\
&&+  r \big(-3 \partial^{a}{\partial^{b}{f^{(4)}_{a b}}}+ \Box{f^{(4)}}+\frac{1}{12} \Box^2{{f^{(2)}}} - \frac{1}{12} {\partial^{a}{\partial^{b}{\Box f^{(2)}_{a b}}}}
\big) +O( r^{\frac{3}{2}})=0\,.
\end{eqnarray}
It is noteworthy that in performing the above computation for the $ r r$ component, as well as those for $ij$ and $i r$ components, we have retained coefficients of the FG expansion up to the $f^{(6)}_{i j}$ terms, so that the constraint equations that we find between different FG coefficients are completely reliable (note that $f^{(5)}_{i j}$ and $f^{(6)}_{i j}$ do not appear in our equations). Actually, a simple dimensional analysis reveals that $f^{(n\geq7)}_{i j}$ terms do not make any contribution to the results of (\ref{f04rr}) to (\ref{f04ir}).

Then, the $ij$ component turns out to be
\begin{eqnarray}\label{f04ij}
&& r \big(- 24 f^{(4)}_{i j} + 2 \partial^{a}{\partial_{i}{f^{(2)}_{a j}}}
+ 2 \partial^{a}{\partial_{j}{f^{(2)}_{a i}}} - \frac{2}{3} \partial_{i}{\partial_{j}{f^{(2)}}}  - 4 \Box{f^{(2)}_{i j}}\nn \\
&&- \frac{2}{3} \partial_{i}{\partial_{j}{\partial^{a}{\partial^{b}{f^{(0)}_{a b}}}}}
 + {\partial^{b}{\partial_{i}{\Box f^{(0)}_{b j}}}} + {\partial^{b}{\partial_{j}{\Box f^{(0)}_{b i}}}}
- \frac{1}{3} {\partial_{i}{\partial_{j}{\Box f^{(0)}}}} - \Box^2{{f^{(0)}_{i j}}}\nn \\
&&+ \eta_{i j} (8 f^{(4)} + \frac{4}{3} \Box{f^{(2)}}
 - \frac{2}{3} \partial^{a}{\partial^{b}{f^{(2)}_{a b}}} - \frac{1}{3} \Box{\partial^{a}{\partial^{b}{f^{(0)}_{a b}}}} + \frac{1}{3} \Box^2{{f^{(0)}}}) \big)\!+\!O( r^{\frac{3}{2}})=0.
\end{eqnarray}
For $i r$ component we have
\begin{eqnarray}\label{f04ir}
&& r^{\frac{1}{2}} \big(- \frac{1}{2} \partial_{i}{f^{(3)}} + \frac{3}{2} \partial^{j}{f^{(3)}_{i j}}- \frac{1}{12} {{\partial_{i}{\Box f^{(1)}}}}
 - \frac{1}{6} \partial_{i}{\partial^{j}{\partial^{a}{f^{(1)}_{j a}}}} + \frac{1}{4} {{\partial^{a}{\Box f^{(1)}_{i a}}}} \big)\nn \\
 &&+  r \big(\!-\! 2 \partial_{i}{f^{(4)}} \!+\! 6 \partial^{j}{f^{(4)}_{i j}}\!-\!\frac{1}{6}  {{\partial_{i}{\Box f^{(2)}}}} \!-\! \frac{1}{3} \partial_{i}{\partial^{j}{\partial^{a}{f^{(2)}_{j a}}}}+ \frac{1}{2} {{\partial^{a}{\Box f^{(2)}_{i a}}}} \big)\!+\!O( r^{\frac{3}{2}})=0.
\end{eqnarray}
As is readily seen, the first line of the above equation gives us a constraint relation between derivatives of $f^{(3)}_{i j}$ and $f^{(1)}_{i j}$. We note that this relation is the same as what one would obtain by linearizing the equation (27) in \cite{Grumiller:2013mxa}.

To check the above equations one can choose another alternative way which we have presented in the appendix B. By inserting $h_{ij}=\eta_{ij}+f_{ij}$ together with the FG expansion of (\ref{fexpand}) into the equations (\ref{eomrr1}), (\ref{eomij1}) and (\ref{eomir1}) we obtain the equations (\ref{f04rr}) to (\ref{f04ir}) exactly.

Now, what we observe by looking at the equations (\ref{f04rr}) to (\ref{f04ir}) is that, none of the FG coefficients $f^{(0)}_{i j}$ to $f^{(3)}_{i j}$
can be completely determined in terms of the other three coefficients. Based on what we know from the holographic renormalization studies, equation (\ref{f04ij}) is not relevant to our analysis for identification of source and response functions. That is due to the presence of the $f^{(4)}_{i j}$ term in this equation, which evidently prevents us from extracting a constraint equation relating only $f^{(0)}_{i j}$ to $f^{(3)}_{i j}$. 

On the other hand from (\ref{f04rr}) and (\ref{f04ir}) and up to the $ r^{\frac{1}{2}}$ order, one finds at most some relations between the divergences and traces of $f^{(2)}_{i j}$ and $f^{(3)}_{i j}$. In other words, these coefficients are not completely determined in terms of some given $f^{(0)}_{i j}$ and $f^{(1)}_{i j}$ functions.

This result is a well known notion within the survey of holographic renormalization; that is we already know that by making just an asymptotic analysis, the most information one can obtain about the response function is its trace and divergence and not more. Then to find further information about it, we need to extend our knowledge by exploring the behavior of the solution inside the bulk \cite{Skenderis,deHaro}.

All in all, the analysis we made in this section leads us to the following important conclusion that is in complete agreement with the results of \cite{Grumiller:2013mxa}:
``The expansion coefficients $f^{(0)}_{i j}$ and $f^{(1)}_{i j}$ can be interpreted as two sources for the dual operators in the corresponding dual quantum field theory, and their responses are somehow related to $f^{(3)}_{i j}$ and $f^{(2)}_{i j}$ respectively, as computed in \cite{Grumiller:2013mxa}."

In the next section we return to the above mentioned "inside the bulk survey" to find the functionality of the response functions in terms of the sources.

%%%%%%%%%%%%%%%%%%%%%%%%%%%%%%%%%%%%%%%%%%%%%%%%%%%%%%%%%%%%%%%%%%%%%%%%%%%%%
\section{Two-point functions}
In order to find the two-point functions, one should calculate the second variation of the on-shell action with respect to the corresponding sources. Alternatively one can obtain it by varying the one-point function with respect to the source. That can be typically stated as
\be
\left\langle O_{1}O_{2}\right\rangle=\frac{i}{\sqrt{-h_{(0)}}}\frac{\delta{\left\langle O_{1}\right\rangle}}{\delta{J_{2}}}\,,
\ee
where $J_{2}$ is the source for the operator $O_{2}$.
So, having the EM tensor, $\tau_{ij}$ and PMR, $P_{ij}$ (clearly defined in the following subsections) as the operators we have to deal with in CG, the two-point functions we are interested in can be written as follows \cite{Skenderis:2009nt}
\bea\label{2pims1}
&&\left\langle P_{i j}(x) P_{k l}(0)\right\rangle=\frac{i}{\sqrt{-h_{(0)}}}\frac{\delta{\left\langle P_{i j}(x)\right\rangle}}{\delta{h_{(1)}^{k l}(0)}}\,,\qquad
\label{2pims2}
\left\langle \tau_{i j}(x) \tau_{k l}(0)\right\rangle=\frac{i}{\sqrt{-h_{(0)}}}\frac{\delta{\left\langle \tau_{i j}(x)\right\rangle}}{\delta{h_{(0)}^{k l}(0)}}\,,\nn\\
\label{2pims3}
&&\qquad\qquad\left\langle P_{i j}(x) \tau_{k l}(0)\right\rangle=\frac{i}{\sqrt{-h_{(0)}}}\frac{\delta{\left\langle P_{i j}(x)\right\rangle}}{\delta{h_{(0)}^{k l}(0)}}=\frac{i}{\sqrt{-h_{(0)}}}\frac{\delta{\left\langle \tau_{k l}(0)\right\rangle}}{\delta{h_{(1)}^{i j}(x)}}\,,
\eea
where $h_{i j}$ is explicitly introduced in (\ref{exp}) and  $\sqrt{-h_{(0)}}=1$  in our computations.

So to perform the above variations we need to know the one-point functions in terms of their sources. This is where the information from inside the bulk comes into the play. That is, we should solve exactly the equations of motion by specifying convenient boundary conditions at the horizon. This is what we are going to do in the upcoming two subsections. In this way, the functionality of the responses are completely determined in terms of the source functions and the desired two-point functions can be computed.

Hence, as we will see, one needs to know the functional relations between $f^{(3)}_{i j}$ and $f^{(2)}_{i j}$ on one side, and
 $f^{(0)}_{i j}$ and $f^{(1)}_{i j}$ on the other side. It is important to note that, because our study is being restricted to the computation of two-point functions, it is sufficient for us to deal with the linearized form of the equations of motion. That means, one should solve the differential equations (\ref{l3}) to (\ref{l2}) exactly.

The equations of motion (\ref{l3}) to (\ref{l2}) are looking complicated to solve. However, as we have shown in the appendix C, by using the reparametrization invariance and the Weyl symmetry, one can write a simple form for the linearized equations. In fact by using the transverse-traceless gauge condition one can show that the linearized equation of motion for metric fluctuation $\textbf{g}_{\mu \nu}$ simplifies as 
	\bea\label{CG.E.O.M}
	&&\frac32 (\Box-\frac{4}{3}\Lambda)(\Box-\frac{2}{3}\Lambda)\textbf{g}_{\mu \nu}=0\,,
	\eea
	which is compatible with what has already been mentioned in \cite{Lu:2011ks,Lu:2011zk}. 
In this way, solutions are classified into two different modes with the following differential equation
\be\label{eommods}
(\Box+ a^{2}) \textbf{g}_{\mu \nu} =0\,,\qquad a^2=2\,,4\,.
\ee
The Einstein modes correspond to $a^2=2$ and the ghost modes to $a^2=4$. To solve (\ref{eommods}), it will be more proper to introduce $\textbf{g}_{\mu \nu}$  through the definition of asymptotically $AdS$ space-time metric in the Poincare coordinates
	\begin{eqnarray}
	ds^{2}_{AAdS}=\frac{dr^{2}}{4 r^{2}} + \frac{1}{r} \eta_{i j} dx^{i} dx^{j}+\textbf{g}_{\mu \nu}dx^{\mu} dx^{\nu}\,.
	\end{eqnarray}
Then the linearized equation (\ref{eommods}) in terms of different components,	$rr$, $ir$ and $ij$ can be written as
		\begin{eqnarray}\label{rrc}
&&4 r^{2} \textbf{g}''_{r r} +14 r \textbf{g}'_{r r}+r\Box{\textbf{g}_{r r}} + \big(a^{2}-2\big) \textbf{g}_{r r}  +2 \partial^{i}{\textbf{g}_{i r}} + \frac{1}{2r} {\textbf{g}^{i}}_{i} =0\,,\nn\\ 
		\label{irc}
&&4 r^{2} \textbf{g}''_{i r} + 10 r \textbf{g}'_{i r} + r \Box{\textbf{g}_{i r}} + \big(a^{2}-6\big) \textbf{g}_{i r}  - 4 r \partial_{i}{\textbf{g}_{r r}} + \partial^{j}{\textbf{g}_{i j}}=0\,,\nn\\ 
		\label{ijc}
&& 4 r^{2} \textbf{g}''_{i j} + 6 r \textbf{g}'_{i j} + r \Box{\textbf{g}_{i j}}  + \big(a^{2}-4\big) \textbf{g}_{i j} - 4 r \partial_{i}{\textbf{g}_{r j}} - 4 r \partial_{j}{\textbf{g}_{r i}} + 8 r \eta_{i j} \textbf{g}_{r r}=0\,,
		\end{eqnarray} 
where prime denotes the derivative with respect to the $r$ coordinate and $\Box=\eta^{i j} \partial_{i}{\partial_{j}}$.
  Based on the fact that these equations are written in the transverse-traceless gauge, one finds the following differential relations between different components
\begin{eqnarray}\label{41}
&&\partial^{i}{\textbf{g}_{i r}}=-4 r \textbf{g}'_{r r}\,,\qquad 
\partial^{j}{\textbf{g}_{i j}}=2 \textbf{g}_{r i}-4 r \textbf{g}'_{r i}\,,\qquad
{\textbf{g}^{i}}_{i}=-4 r \textbf{g}_{r r}\,.
\end{eqnarray}
In order to find the solutions, we consider $\textbf{g}_{\mu \nu}(r,x^{i})=e^{-ip.x} {\textbf{g}}_{\mu \nu}(r)$ as an ansatz for metric fluctuations. Then depending on how to consider the three-momentum $p_{i}$, as a time-like, space-like or light-like vector, one finds  different solutions for ${\textbf{g}}_{\mu \nu}(r)$.
%%%%%%%%%%%%%%%%%%%%%%%%%%%%%%%%%%%%%%%%%%%%%%%%%%%%%%%%%%%%%%%%%%%%%%%
\subsection{Time-like modes}
To construct a time-like mode it is possible to choose $p_{i}=E\delta^{t}_{i}$ as a time-like three-momentum. By using the gauge conditions in (\ref{41}), the differential equations (\ref{rrc}) simplify as follows
	\begin{eqnarray}\label{set1}
&&	4 r^{2} \textbf{g}''_{r r} + 6 r \textbf{g}'_{r r} +(r E^2+ a^{2} -4) \textbf{g}_{r r} =0\,,\nn\\
&&	4 r^{2} \textbf{g}''_{i r} + 6 r \textbf{g}'_{i r} +(r E^2+ a^{2} -4) \textbf{g}_{i r}- 4 r \partial_{i}{\textbf{g}_{r r}} = 0\,,\nn\\
&&	 4 r^{2} \textbf{g}''_{i j} + 6 r \textbf{g}'_{i j} +(r E^2+ a^{2} -4) \textbf{g}_{i j} - 4 r \partial_{i}{\textbf{g}_{r j}} - 4 r \partial_{j}{\textbf{g}_{r i}} + 8 r \eta_{i j} \textbf{g}_{r r}=0\,.
	\end{eqnarray}
The solution for components $\textbf{g}_{r r}$, $\textbf{g}_{x r}$, $\textbf{g}_{y r}$ and $\textbf{g}_{x y}$ can be obtained easily by solving the differential equations in (\ref{set1}) directly.
For other components, we need to use the gauge conditions (\ref{41}) in the following form 
	\begin{eqnarray}\label{gauge1}
	&&\textbf{g}_{t r} =\frac{4i}{E} r \textbf{g}'_{r r}\,, \quad \textbf{g}_{t i} =\frac{2i}{E} (2r \textbf{g}'_{r i} -\textbf{g}_{r i})\,,\quad \textbf{g}_{x x} +\textbf{g}_{y y} = \textbf{g}_{t t} -4r \textbf{g}_{r r}\,.
	\end{eqnarray}
Since we are interested to find the solutions in the Gaussian (Fefferman-Graham) gauge (solutions for equations of motion (\ref{l3}) to (\ref{l2})), we need also to know the gauge transformation parameters, $\xi^{\mu}(r,x^{i})=e^{-iEt} {\xi}^{\mu}(r)$.	The relation between Gaussian modes and transverse-traceless modes can be found as 
\be\label{gg}
r^{-1}f_{\mu \nu}=\tb{g}_{\mu \nu}+\nabla_{\mu} \xi_{\nu}+ \nabla_{\nu} \xi_{\mu}\,,
\ee
and the Gaussian gauge is defined through the 
$ f_{r r} = f_{r i} = 0 $, 
	therefore it follows that the different components of $\xi_{\mu}$ should satisfy the following differential equations
	\begin{eqnarray}\label{repar1}
	(r \xi_{r})' =-\frac{r}{2} \textbf{g}_{r r}\,,\quad
	(r \xi_{t})' =-r \textbf{g}_{r t} + irE \xi_{r}\,,\quad
	(r \xi_{x})' =-r \textbf{g}_{r x} \quad 
	(r \xi_{y})' =-r \textbf{g}_{r y} \,.
	\end{eqnarray}
%%%%%%%%%%%%%%%%%%%%%%%%%%%%%%%%%%%%%%%%%%%%%%%%%%%%%%%%%%%%%%%%%%%%
\subsubsection{Ghost modes}
First let us discuss the ghost modes with $a^{2}=4$. To solve equations in (\ref{set1}) we need to consider the ``infalling boundary condition in the Bulk".
That means that we require the behavior of solution around the Poincare horizon of the bulk, i.e. at $r\rightarrow \infty$, to be of the form $e^{\pm iE(t-\sqrt{r})}$, so that the fluctuation modes go toward the horizon as time passes and do not come out of it,\footnote{For a detailed discussion about different boundary conditions for Euclidean and Minkowski signature see for example \cite{Son:2002sd}.}

Now by adjusting these solutions such that they satisfy the gauge conditions in (\ref{gauge1}) we will obtain the form of fluctuations in the transverse-traceless gauge. Then we must substitute our solutions into (\ref{repar1}). In this way we can find the gauge transformation parameters $\xi_{\mu}$ required for going  to the Gaussian gauge.
By using (\ref{gg})
we finally find the following solutions in the Gaussian gauge
	\begin{eqnarray}
&& f_{t t} =c_{1} \frac {16 i}{E^4} e^{i E \sqrt{r}}+ (4 +2 E^2 r )c_{2}-2 i E c_{3}\,,\qquad
f_{x y}=c_{9} \sqrt{r} e^{i E \sqrt{r}}\,,\nn\\
&& f_{x t}= -c_{4} \frac {8}{E^3} e^{i E \sqrt{r}}-i E c_{5}\,,\qquad
 f_{y t}= -c_{6} \frac {8}{E^3} e^{i E \sqrt{r}}-i E c_{7}\,,\nn\\
&& f_{x x}= \frac {1}{E^4} e^{i E \sqrt{r}} \big(8 (2 i+E \sqrt{r}) c_{1}+E^4 \sqrt{r} c_{8}\big)-4 c_{2}\,,\nn\\
&& f_{y y}= \frac {1}{E^4} e^{i E \sqrt{r}} \big(8 (2 i+E \sqrt{r}) c_{1}-E^4 \sqrt{r} c_{8}\big)-4 c_{2}\,,
	\end{eqnarray}
where each $c_{n}$ is a constant of  integration. As a double check we can verify that the above solutions satisfy (\ref{l3}) to (\ref{l2}). 

These solutions will give rise to five different ghost modes if one chooses a specific combination of the constants $c_{n}$'s so that, the expansion of these modes near the boundary at $r=0$ contains no $r^{0}$ terms. We impose this condition in order to make sure that while computing the two-point functions by varying one-point functions with respect to the source $f^{(1)}_{i j}$, the source $f^{(0)}_{i j}$ is automatically turned off in the last step of calculation. After doing this we can rewrite  the following matrix representation for the ghost modes
\bea\label{tlike}
&&f^{G(1)}_{i j} =i (1-e^{i E \sqrt{r}}) \mathcal{M}_1\,,\qquad
f^{G(2)}_{i j} =i (1-e^{i E \sqrt{r}}) \mathcal{M}_2\,,\nn\\
&&f^{G(3)}_{i j} =\sqrt{r} e^{i E \sqrt{r}} \mathcal{M}_3\,,\qquad\quad\,\,\,
	f^{G(4)}_{i j} =\sqrt{r} e^{i E \sqrt{r}} \mathcal{M}_4\,,\nn \\
&& f^{G(5)}_{i j} =\big(\frac{2 i}{E} (1-e^{i E \sqrt{r}})-\sqrt{r} e^{i E \sqrt{r}}\big)\mathcal{I}+\big(\sqrt{r} e^{i E \sqrt{r}}-iEr\big) \mathcal{M}_5\,,
\eea
where
\begin{eqnarray}
&	\mathcal{M}_1=
	\begin{pmatrix}
	0 & 1 & 0 \\
	1 & 0 & 0 \\
	0 & 0 & 0 \\
	\end{pmatrix}\,,\quad 
	\mathcal{M}_2 =
	\begin{pmatrix}
	0 & 0 & 1 \\
	0 & 0 & 0 \\
	1 & 0 & 0 \\
	\end{pmatrix}\,, \quad
 \mathcal{M}_3 =
	\begin{pmatrix}
	0 & 0 & 0 \\
	0 & 1 & 0 \\
	0 & 0 & -1 \\
	\end{pmatrix}\,,\nn \\
&	\mathcal{M}_4 =
	\begin{pmatrix}
	0 & 0 & 0 \\
	0 & 0 & 1 \\
	0 & 1 & 0 \\
	\end{pmatrix} \,,\quad
	\mathcal{M}_5 =
		  \begin{pmatrix}
	    1&0&0\\
	    0&0&0\\
	    0&0&0
	  \end{pmatrix}\,,\quad
		\mathcal{M}_6 =
		\begin{pmatrix}
	    -1&0&0\\
	    0&0&0\\
	    0&0&1
	  \end{pmatrix}\,.
	\end{eqnarray}
	It is worthwhile to mention that in writing the above equation we have profited the symmetry of equations of motion under the $SO(2)$ subgroup of the symmetry group of the boundary theory and adjusted $c_{n}$'s in such a way that this symmetry is manifest in solutions as well. Here (\ref{tlike}) are written in vector, tensor and scalar representations of $SO(2)$ respectively. 
	Now we are able to write  the covariant time-like solution in terms of three-momentum $p_{i}$ as follows 
	\begin{eqnarray}\label{4.52}
&&	f^{G(1,2)}_{i j} =i (1-e^{i {\left|p\right|} \sqrt{r}}) \frac{1}{\left|p\right|}(p_{i} \epsilon^{1,2}_{j}+p_{j} \epsilon^{1,2}_{i})\,, \qquad 	\label{4.53}
	f^{G(3,4)}_{i j} =\sqrt{r} e^{i {\left|p\right|} \sqrt{r}} M^{1,2}_{i j}\,, \nn \\
&&	\label{4.54}
		f^{G(5)}_{i j} =\frac{2 i}{{\left|p\right|}} (1-e^{i {\left|p\right|} \sqrt{r}}) (\eta_{i j}-2\frac{p_{i} p_{j}}{p^2})
	-\sqrt{r} e^{i {\left|p\right|} \sqrt{r}} (\eta_{i j}-\frac{p_{i} p_{j}}{p^2})
	   +i {\left|p\right|} r \frac{p_{i} p_{j}}{p^2}\,,
	  \end{eqnarray}
where $\epsilon^{1,2}_{i}$ are two transverse polarizations from which, $M^{1,2}_{i j}$ are constructed 
\begin{eqnarray}
M^{1}_{i j}=\epsilon^{1}_{i} \epsilon^{1}_{j}-\epsilon^{2}_{i} \epsilon^{2}_{j}\,,\qquad M^{2}_{i j}=\epsilon^{1}_{i} \epsilon^{2}_{j}+\epsilon^{2}_{i} \epsilon^{1}_{j}\,.
\end{eqnarray}
It is easy to verify that the only Lorentz covariant form of the solutions that also transforms covariantly under $SO(2)$ is  (\ref{4.52}).
	More precisely, $p_{i}$ and $\eta_{i j}$ are scalar objects under $SO(2)$, while $\epsilon^{1,2}_{i}$ are two vectors from which $\sum{A_{IJ}\epsilon^{I}_{i}\epsilon^{J}_{j}}$ tensors can be constructed.

 Having the solutions (\ref{4.52}), we are able to calculate the two-point function for the dual QFT operator whose source is $f^{(1)}_{i j}$. To this end we will follow the same approach as in \cite{Johansson:2012fs}.
%%%%%%%%%%%%%%%%%%%%%%%%%%%%%%%%%%%%%%%%%%%%%%%%%%%%%%%%%%%%%%%%%%%%%%%%%%%%%%%%%%%%%%%%%%%
 \subsubsection{Einstein modes}
As we mentioned, this mode is specified by $a^{2}=2$. By doing the same steps as what we have done for the ghost modes we can find the following solutions in the Gaussian gauge 
	\begin{eqnarray}
	&&f_{t t} = 2c_{1} (2 + E^{2} r)- 2c_{2}iE\,, \nn\\
	&&f_{t x} = -iE c_{3}\,, \qquad
	f_{t y} = -iE c_{4}\,, \nn\\
	&&f_{x x} = -4 c_{1} +
	    c_{5}(1 - iE \sqrt{r})e^{i E \sqrt{r}}\,, \nn\\
	&&f_{y y} = -4 c_{1} -
	    c_{5}(1 - iE \sqrt{r})e^{i E \sqrt{r}}\,, \nn\\
	&&f_{x y} = -3ic_{6}  \frac{1 -i E \sqrt{r}}{2 E^{2}}e^{iE \sqrt{r}}\,.
	\end{eqnarray}
We will obtain six Einstein modes if we choose a specific combination of constants $c_{n}$'s. Note that all $r^{\frac12}$ terms are absent here automatically. Then these modes take the following matrix representation 
\bea\label{4.60A}
&&f^{E(1)}_{i j} = E \mathcal{M}_1\,,\qquad
f^{E(2)}_{i j} = E \mathcal{M}_2\,, \qquad
 f^{E(3)}_{i j} =e^{iE \sqrt{r}}(1-iE\sqrt{r}) \mathcal{M}_3\,, \nn\\
&& f^{E(4)}_{i j} =e^{iE \sqrt{r}}(1-iE\sqrt{r}) \mathcal{M}_4\,,\qquad
f^{E(5)}_{i j} =E^{2} \mathcal{M}_5\,,\nn\\
&& f^{E(6)}_{i j} =(4+E^2 r) \mathcal{M}_5 -2 \mathcal{I}\,.
\eea
	Similarly we can write  the most general time-like solution in terms of $p_{i}$ as
	\begin{eqnarray}\label{4.60}
&&	f^{E(1,2)}_{i j} =(p_{i} \epsilon^{1,2}_{j}+p_{j} \epsilon^{1,2}_{i}) \,,\nn \\
&&	\label{4.61}
	f^{E(3,4)}_{i j} =e^{i |p| \sqrt{r}} (1-i|p| \sqrt{r}) M^{1,2}_{i j} \,,\nn \\
&&	\label{4.62}
	f^{E(5)}_{i j} =p_{i} p_{j} \,,\quad	\label{4.63}
	f^{E(6)}_{i j} =-2\eta_{i j}+ r p_{i} p_{j}\,.
	\end{eqnarray}
%%%%%%%%%%%%%%%%%%%%%%%%%%%%%%%%%%%%%%%%%%%%%%%%%%%%%%%%%%%%%%%%%%%%%%%%%%%%%
 \subsubsection{Two-point functions in momentum space}
After finding the explicit form of solutions in three-momentum space, now we can extract the source and response functions and perform the required variations to obtain the two-point functions.
As explained earlier, to do this calculation first of all we need the one-point functions of the operators $\tau_{i j}$ and $P_{i j}$. These have been figured out in \cite{Grumiller:2013mxa} and we are going to use the results. As we will see, in this way we can find the two-point functions in (\ref{2pims1}) in momentum space and after that we just need to do some Fourier transformations to find their form in the configuration space.

So let us start to calculate the two-point function of PMR tensor in momentum space, then in the next sections we will repeat this analysis for other relevant two-point functions.

 For later convenience, it is advantageous to explicitly write the expansions of $h^{(0)}_{i j}$, $h^{(1)}_{i j}$, $h_{(0)}^{i j}$ and $h_{(1)}^{i j}$.
	Considering the Einstein modes, we find the following relations (note that we should have $h^{(0)}_{i j}h_{(0)}^{j m}=\eta_{i}^{m}$)
	\begin{equation}
h^{(0)}_{i j}=\eta_{i j}+e^{ip.x} \sum_{I=1}^{6} B_{I} t^{I}_{i j}\,,\qquad
h_{(0)}^{i j}=\eta^{i j}-e^{ip.x} \sum_{I=1}^{6} B_{I} t_{I}^{i j}\,,
	\end{equation}
where $t^{I}_{i j}$ are some symmetric basis in terms of which we will determine the source functions $f^{(0)}_{i j}$. We will explicitly introduce this basis in the subsequent sections.
	On the other hand according to our convention $h_{(0)}^{i j}=\eta^{i j}+f_{(0)}^{i j}$, we find that $f_{(0)}^{i j}$ can be written as
	\begin{equation}\label{4.123}
	f_{(0)}^{i j}=-e^{ip.x} \sum_{I=1}^{6} B_{I} t_{I}^{i j}\,.
	\end{equation}
In a similar way if we consider the ghost modes, we will find the following form for $h^{(1)}_{i j}$, which up to the linear order in fluctuations is the same as $f^{(1)}_{i j}$ 
	\begin{equation}\label{4.122}
	f^{(1)}_{i j}=e^{ip.x} \sum_{I=1}^{5} A_{I} e^{I}_{i j}\,,\qquad
	f_{(1)}^{i j}=e^{ip.x}\sum_{I=1}^{5} A^{I} e_{I}^{i j}\,,\qquad e_{I}^{i j}=\eta^{i a} \eta^{j b} e^{I}_{a b}\,.
	\end{equation}
%	where we have used the fact that $h_{i j}h^{j m}=\eta_{i}^{m}$, which leads to $h_{(1)}^{i %j}=h_{(0)}^{i m}h_{(0)}^{j n}h^{(1)}_{m n}$.
In \cite{Grumiller:2013mxa}, it has been argued that $f^{(1)}_{i j}$ may be consistently considered as a source for the Partial Massless Response (PMR) operator, denoted by $P_{i j}$. So to compute the two-point function of PMR one must calculate the variation of $P_{i j}$ with respect to $f^{(1)}_{i j}$. The value of $P_{i j}$ is given in \cite{Grumiller:2013mxa} as \footnote{From now on we consider $\ell$, the radius of AdS space, to be equal to 1 and we choose $\sigma=-1$.}
 \bea\label{Pij}
P_{ij}=-\tfrac{4\,\sigma}{\ell}\,E^{(2)}_{ij}\,,\qquad\label{70}
E^{(2)}_{ij} =  - \tfrac{1}{2\ell^2} \psi_{ij}^{\SO}+ \tfrac{\sigma}{2}\, \big(R_{ij}^{\LO} - \tfrac13 h_{ij}^{\LO}R^{\LO}\big) + \tfrac{1}{8\ell^2} h^{\FO} \psi_{ij}^{\FO}\,, \label{eq:CG23} 
 \eea
where $\psi^{(n)}_{i j}$ is defined as the traceless part of $h^{(n)}_{i j}$
	\begin{eqnarray}
\psi^{(n)}_{i j}=h^{(n)}_{i j}-\frac{1}{3}h^{(0)}_{i j}h^{(n)}\,,\qquad h^{(n)}=h_{(0)}^{i j}h^{(n)}_{i j}\,.
	\end{eqnarray} 
To write $f^{(1)}_{i j}$, we introduce a set of five symmetric matrices as our basis \footnote{It is important to note that all over in this section, we drop the coefficients  $e^{ip.x}$ from both source and response functions while computing the two-point functions. This is due to the fact that we are doing this calculation in the momentum space. At the end we will do a Fourier transformation to find the form of correlation functions in the configuration space.}. 
In order to do that, we extract the coefficients of $r^\frac12$ in the $r$-expansions of the five ghost modes given in (\ref{4.52}). The results are as follows
	\begin{eqnarray}\label{e.p}
	e^{1,2}_{i j} =(p_{i} \epsilon^{1,2}_{j}+p_{j} \epsilon^{1,2}_{i})\,,\qquad
	e^{3,4}_{i j} = M^{1,2}_{i j} \,,\qquad
	e^{5}_{i j} =(\eta_{i j}-3\frac{p_{i} p_{j}}{p^2})\,.
	\end{eqnarray}     
Since a generic ghost solution can be written as $f^{G}_{i j}=\sum{A_{I} f^{G{I}}_{i j}}$, with the summation being over different ghost modes, it is clear that the  coefficient of $r^\frac12$ in the Fefferman-Graham  expansion of this mode can be presented as $f^{G(1)}_{i j}=\sum{A_{I} e^{I}_{i j}}\,$. then, in this formulation we should find an expression for $P_{i j}$ in terms of the basis $e^{I}_{i j}$.

It is important to note that in all subsequent calculations, one only needs to consider contribution from those terms in the one-point functions that are linear in metric fluctuations. In fact, higher order terms after turning off the sources at the end of two-point function evaluation, or equivalently after evaluating the first variation of PMR in the $AdS$ background metric, will vanish and so they do not make any contribution.

Regarding the above consideration as well as the fact that $f^{(0)}_{i j}$ and $f^{(1)}_{i j}$ are supposed to be two independent sources, we remain just with the following form for PMR from (\ref{eq:CG23})
	\begin{eqnarray}\label{pij.pp}
	P_{i j}=-2\psi^{(2)}_{i j}\,.
	\end{eqnarray}
	
Note that in the above equation we retained just the part that will contribute to the correlator of two PMR's. As we will see in section (\ref{316}), if we are interested to find the correlator of PMR with energy-momentum tensor, we must consider all terms of $E^{(2)}_{ij}$ except the last one in (\ref{Pij}).
Also as a result of the aforementioned consideration it is clear that the only relevant part of $\psi^{(n)}_{i j}$ would be
	\begin{eqnarray}\label{psi.n}
	\psi^{(n)}_{i j}|_{relv}.=f^{(n)}_{i j}-\frac{1}{3}\eta_{i j}f^{(n)}\,,\qquad f^{(n)}=\eta^{a b}f^{(n)}_{a b}\,.
	\end{eqnarray}
Therefore to obtain the two-point correlation function of PMR, we need to compute $f^{(2)}_{i j}$ from the ghost modes. Since the Einstein modes do not contribute to $f^{(1)}_{i j}$, they do not enter into the calculation of $\frac{\delta{f^{(2)}_{i j}}}{\delta{f_{(1)}^{k l}}}$.

 To this end, we are now going to write the coefficient of $r$  in the $r$-expansion of the ghost modes, in terms of our 5-fold basis. We find that the fifth mode does not contribute to the $r$ coefficient and the contribution from the other modes gives the following form for $f^{G(2)}_{i j}$
	\begin{eqnarray}\label{f2G}
	f^{G(2)}_{i j}=\frac{i {\left|p\right|}}{2} A_{1} e^{1}_{i j}+\frac{i {\left|p\right|}}{2} A_{2} e^{2}_{i j}+i {\left|p\right|} A_{3} e^{3}_{i j}+i {\left|p\right|} A_{4} e^{4}_{i j}\,.
	\end{eqnarray}
	So in order to compute the first variation of $f^{(2)}_{i j}$ with respect to $f^{(1)}_{i j}$, we need some relations between $A_{1}$ to $A_{4}$ and $f^{(1)}_{i j}$.	We use the following transverse relations
	\begin{eqnarray}\label{4.80}
	e^{K}_{i j} e_{I}^{i j}=-2 {\left|p\right|}^{2} \delta^{K}_{I}\,; \,\,K=1,2, \qquad
	e^{K}_{i j} e_{I}^{i j}=2 \delta^{K}_{I}\,; \,\, K=3,4, \qquad
	e^{5}_{i j} e_{I}^{i j}=6\delta^{5}_{I}\,,
	\end{eqnarray}
	from which and by considering (\ref{4.122}) it is readily seen that (${\left|p\right|}^{2}=-p^{2}$)
	\begin{eqnarray}\label{4.81}
	e^{K}_{i j} f_{(1)}^{i j}=2 p^{2} A_{K}\,; \,\, K=1,2,\qquad
	e^{K}_{i j} f_{(1)}^{i j}=2 A_{K}\,; \,\, K=3,4\,, \qquad
	e^{5}_{i j} f_{(1)}^{i j}=6 A_{5}\,.
	\end{eqnarray}
	 Consequently we find that 
	\begin{eqnarray}\label{si2f1}
	&&\frac{\delta{\psi^{(2)}_{i j}}}{\delta{f_{(1)}^{k l}}}=\frac{\delta{f^{(2)}_{i j}}}{\delta{f_{(1)}^{k l}}}-\frac{1}{3}\eta_{i j}\eta^{m n}\frac{\delta{f^{(2)}_{m n}}}{\delta{f_{(1)}^{k l}}}=\frac{\delta{f^{(2)}_{i j}}}{\delta{f_{(1)}^{k l}}} \nn\\
	&&\,\,\,\,\,\,\,\,\,\,\,\,\,\,\,=-\frac{i}{4 {\left|p\right|}}(e^{1}_{i j} e^{1}_{k l}+e^{2}_{i j} e^{2}_{k l})+\frac{i {\left|p\right|}}{2} (e^{3}_{i j} e^{3}_{k l}+e^{4}_{i j} e^{4}_{k l})\,,
	\end{eqnarray}
	where the second equality comes from the traceless of $e^{(1)}_{i j}$ to $e^{(4)}_{i j}$.
	Now if we write the $e^{I}_{i j}$'s in the above formula in terms of the polarization vectors (\ref{e.p}) and use the following relation 
	\begin{eqnarray}
	\eta_{i j}=\frac{p_{i} p_{j}}{p^2}+\epsilon^{1}_{i} \epsilon^{1}_{j}+\epsilon^{2}_{i} \epsilon^{2}_{j}\,,
	\end{eqnarray}
	we will find that
	\begin{eqnarray}\label{330}
	&&e^{1}_{i j} e^{1}_{k l}+e^{2}_{i j} e^{2}_{k l}=-\frac{1}{{\left|p\right|}^2}(p_{i} p_{k} \Theta_{j l}+p_{i} p_{l} \Theta_{j k}+p_{j} p_{k} \Theta_{i l} +p_{j} p_{l} \Theta_{i k})\,,\nn \\
	&&e^{3}_{i j} e^{3}_{k l}+e^{4}_{i j} e^{4}_{k l}=\frac{1}{{\left|p\right|}^4}(\Theta_{i k} \Theta_{j l}+\Theta_{i l} \Theta_{j k}-\Theta_{i j} \Theta_{k l})\,,
	\end{eqnarray}
where $\Theta_{i j}$ is defined to be 
\be\label{theta}
\Theta_{i j}=\eta_{i j} p^{2}-p_{i} p_{j}\,.
\ee
	So the two-point function of PMR is given by 
	\begin{eqnarray}\label{2pP}
	&&\langle P_{i j}P_{k l}\rangle = i\frac{\delta{\langle P_{i j}\rangle}}{\delta{f_{(1)}^{k l}}}=-2 i\frac{\delta{f^{(2)}_{i j}}}{\delta{f_{(1)}^{k l}}}=\frac{1}{2 {\left|p\right|}^3}(p_{i} p_{k} \Theta_{j l}+p_{i} p_{l} \Theta_{j k}+p_{j} p_{k} \Theta_{i l} +p_{j} p_{l} \Theta_{i k})\nn\\
	&&\qquad \quad \,\,\,\,+\frac{1}{{\left|p\right|}^3} (\Theta_{i k} \Theta_{j l}+\Theta_{i l} \Theta_{j k}-\Theta_{i j} \Theta_{k l})\,,
	\end{eqnarray}
	where in the last equality we have written (\ref{si2f1}) in terms of $\Theta_{i j}$ by using
	(\ref{330}).
%%%%%%%%%%%%%%%%%%%%%%%%%%%%%%%%%%%%%%%%%%%%%%%%%%%%%%%%%%%%%%%%%%%%%%%%%%
\subsubsection{Two-point function of PMR in configuration space}
 In what follows, we will compute the two-point function of PMR in the configuration space. To do this we just need a Fourier transformation that we are going to perform in the Euclidean space
\begin{eqnarray}\label{pg1cs}
	&&i\frac{\delta{\langle P_{i j}\rangle}}{\delta{f_{(1)}^{k l}}}=\int d^{3}p e^{-ip.x}\Big(\frac{1}{2 {\left|p\right|}^3}(p_{i} p_{k} \Theta_{j l}+p_{i} p_{l} \Theta_{j k}+p_{j} p_{k} \Theta_{i l} +p_{j} p_{l} \Theta_{i k})\nn\\
	&&\qquad \quad \,\,+\frac{1}{{\left|p\right|}^3} (\Theta_{i k} \Theta_{j l}+\Theta_{i l} \Theta_{j k}-\Theta_{i j} \Theta_{k l})\Big)\,.
	\end{eqnarray}
Now using the fact that $\hat{\Theta}_{i j} e^{-ip.x}=\Theta_{i j} e^{-ip.x}$ with $\hat{\Theta}_{i j}$ operator defined by $\hat{\Theta}_{i j}=\partial_{i} \partial_{j}-\eta_{i j}\Box $,  we can write (\ref{pg1cs}) equivalently as
\begin{eqnarray}\label{pf1int}
&&i\frac{\delta{\langle P_{i j}\rangle}}{\delta{f_{(1)}^{k l}}}=\Big(-\frac{1}{2}(\partial_{i} \partial_{k} \hat{\Theta}_{j l}+\partial_{i} \partial_{l} \hat{\Theta}_{j k}+\partial_{j} \partial_{k} \hat{\Theta}_{i l} +\partial_{j} \partial_{l} \hat{\Theta}_{i k})\nn\\
	&&\qquad \quad \,\,+(\hat{\Theta}_{i k} \hat{\Theta}_{j l}+\hat{\Theta}_{i l} \hat{\Theta}_{j k}-\hat{\Theta}_{i j} \hat{\Theta}_{k l})\Big)\int d^{3}p e^{-ip.x}\frac{1}{{\left|p\right|}^3}\,.
\end{eqnarray}
So we should just compute the following integral 
\begin{eqnarray}
-2\pi\int_{0}^{\infty}{d|p|\int_{0}^{\pi}dcos{\theta} e^{-i|p||x|cos{\theta}}(\frac{1}{{\left|p\right|}}) }=
\frac{4\pi}{|x|}\int{dy \frac{1}{y^2}sin{(|x|y)} }\,,
\end{eqnarray}
where $y=|p|$ and this integral diverges.
In order to regulate this integral we change the power of the variable $y$ by some infinitesimal positive amount $\epsilon$ 
\begin{eqnarray}\label{reg}
\frac{4\pi}{|x|}\int{dy \frac{1}{y^{2-\epsilon}}sin{(|x|y}) }\!\!&=&\!\!-4\pi|x|^{-\epsilon}cos{(\frac{\epsilon \pi}{2})}\Gamma[-1+\epsilon]\nn\\
\!\!&=&\!\!\frac{4\pi}{\epsilon}-4\pi(-1+\gamma+\log{|x|})+O(\epsilon)\,.
\end{eqnarray}
So (\ref{pf1int}) after being regularized, turns out to be
\begin{eqnarray}\label{pf1intfinal}
&&\langle P_{i j}P_{k l}\rangle ~=~ i\frac{\delta{\langle P_{i j}\rangle}}{\delta{f_{(1)}^{k l}}}=4\pi\big(\frac{1}{2}(\partial_{i} \partial_{k} \hat{\Theta}_{j l}+\partial_{i} \partial_{l} \hat{\Theta}_{j k}+\partial_{j} \partial_{k} \hat{\Theta}_{i l} +\partial_{j} \partial_{l} \hat{\Theta}_{i k})\nn\\
	&&\qquad\qquad-(\hat{\Theta}_{i k} \hat{\Theta}_{j l}+\hat{\Theta}_{i l} \hat{\Theta}_{j k}-\hat{\Theta}_{i j} \hat{\Theta}_{k l})\big)\log{|x|}\,,
\end{eqnarray}
where we have neglected $-1+\gamma$ in (\ref{reg}) because it does not contribute after being subjected to the derivatives included in $\hat{\Theta}_{i j}$.

Before closing this section, let us analyze the result (\ref{pf1intfinal}) with more details. It is shown in \cite{Grumiller:2013mxa} that the $P_{ij}$ is traceless. According to the ADM decomposition \cite{Arnowitt:1962hi} of a traceless tensor $A_{ij}$ we have
\begin{equation}\label{sossos}
	A_{i j} = \nabla_{i}V_{j}+ \nabla_{j}V_{i}+P^{TT}_{i j} + (\nabla_{i}\nabla_{j}-\frac{1}{3}\eta_{i j}\nabla^{2})S\,,
\end{equation}
with $V_{i}$ being a transverse vector  ($\nabla^{i}V_{i} = 0$), $P^{TT}_{i j}$ a transverse-traceless tensor and $S$ being a scalar. Moreover, according to (\ref{CG.E.O.M}), PMR-modes are actually massive gravitons with $M^2 = \frac{2\Lambda}{3}$. It is discussed in \cite{Deser:2001pe,Deser:2001us} that the massive gravitons with this special amount of mass contain only spin two and spin one parts \footnote{This phenomenon in known as “partial masslessness” \cite{Deser:2001pe}. For more details on PMR modes see \cite{Deser:2012qg, Deser:2013bs}.}. Therefore the York decomposition for $P_{ij}$ becomes
\begin{equation}\label{sos}
	P_{i j} = \nabla_{i}V_{j}+ \nabla_{j}V_{i}+P^{TT}_{i j}.
\end{equation}
To be more precise, substituting (\ref{e.p}) in (\ref{f2G}) and using (\ref{pij.pp}) and (\ref{psi.n}), one can see that in the momentum space
\begin{equation}\label{340}
P_{i j} = -i|p|\Big(p_{i}(A_{1}\epsilon^{1}_{j}+A_{2}\epsilon^{2}_{j})+p_{j}(A_{1}\epsilon^{1}_{i}+A_{2}\epsilon^{2}_{i})\Big)-2i|p|(A_{3}M^{1}_{i j}+A_{4}M^{2}_{i j})\,.
\end{equation}
A comparison with (\ref{sos}) shows that $|p|(A_{1}\epsilon^{1}_{i}+A_{2}\epsilon^{2}_{i})$, plays the role of transverse vector with two degrees of freedom and $|p| M^{1,2}_{i j}$ are two  transverse-traceless tensors. As can be seen from (\ref{340}), the scalar degree of freedom is absent  in conformal gravity which is in agreement with the discussion in \cite{Deser:2001us,Deser:2001pe}\footnote{The field content of CG at the linearized level also found in \cite{Lee:1982cp,Riegert:1984hf}. CG also is formulated as a second order theory \cite{Metsaev:2007fq}. In that formulation the field content can be seen more clearly.}.

Therefore we conclude that the transverse vector part of $P_{i j}$ corresponds to the $e^{1}_{i j}$ and $e^{2}_{i j}$ parts of the metric fluctuation  and the transverse-traceless tensor part to $e^{3}_{i j}$ and $e^{4}_{i j}$.
Moreover by using the orthogonality relations between bases, it is meaningful to compute $\frac{\delta P_{i j}}{\delta f^{k l}_{(1)}}|_{TV,TV}$ and $\frac{\delta P_{i j}}{\delta f^{k l }_{(1)}}|_{TT,TT}$ which corresponds to letting only $A_{1,2}$ and $A_{3,4}$ be non-zero respectively. In view of the above considerations, from (\ref{2pP}) we have
\begin{equation}\label{TVTV}
\frac{\delta P_{i j}}{\delta f_{(1)}^{k l }}|_{TV,TV}=-\frac{i}{2|p|^{3}} p_{(i}\Theta_{j)(l}p_{k)}\,,
\end{equation}
and
\begin{equation}\label{TTTT}
\frac{\delta P_{i j}}{\delta f_{(1)}^{k l }}|_{TT,TT}=-\frac{i}{|p|^{3}} (\Theta_{i k}\Theta_{j l}+\Theta_{i l}\Theta_{j k}-\Theta_{i j}\Theta_{k l})\,.
\end{equation}
To extract the form of two-point function $\langle V_{i}V_{j}\rangle$ from (\ref{TVTV}) we write 
\begin{eqnarray}
&&\langle \partial_{i}V_{j}+\partial_{j}V_{i}| \partial_{k}V_{l}+\partial_{l}V_{k} \rangle \equiv 
-p_{i}p_{k} \langle \widetilde{V}_{j}|\widetilde{V}_{l}\rangle- p_{i}p_{l}\langle \widetilde{V}_{j}|\widetilde{V}_{k}\rangle- p_{j}p_{k}\langle \widetilde{V}_{i}|\widetilde{V}_{l}\rangle- p_{j}p_{l}\langle \widetilde{V}_{i}|\widetilde{V}_{k}\rangle\,,\nonumber\\
&&
\end{eqnarray}
where by $\equiv$ we mean the same expression in the Fourier transformed form and $\widetilde{V}_{i}$ stands for the vector in momentum space.
The above expression is supposed to be the same as the first parenthesis in (\ref{2pP}),
where one can identify 
\begin{equation}
\langle \widetilde{V}_{i}|\widetilde{V}_{j}\rangle ~=~ \frac{-1}{2 |p|^{3}}\Theta_{i j}\,,
\end{equation}
and so on. Therefore in configuration space we have
\begin{equation}\label{VV}
\langle V_{i}(x)|V_{j}(0)\rangle = 2\pi \hat{\Theta}_{i j} log|x|= -4\pi \frac{x_{i} x_{j}}{|x|^{4}}\,.
\end{equation}
Note that in this relation the transverse property of $V_{i}(x)$ ensures the absence of $\frac{\eta_{ij}}{|x|^{2}}$ term in the final result.

Finally from the $TT$ component of the York decomposition we find
\begin{eqnarray}
	&&\langle P^{TT}_{i j}| P^{TT}_{k l}\rangle \,\equiv\,\langle \widetilde{P}^{TT}_{i j}| \widetilde{P}^{TT}_{k l}\rangle\,,
\end{eqnarray}
where it is supposed to be the same as the second parenthesis in (\ref{2pP}), 
therefore in configuration space one obtains
\begin{equation}\label{PTT}
\langle P^{TT}_{i j}| P^{TT}_{k l}\rangle = -4 \pi ( \hat{\Theta}_{i k} \hat{\Theta}_{j l}+ \hat{\Theta}_{i l} \hat{\Theta}_{j k}- \hat{\Theta}_{i j}\hat{\Theta}_{k l}) log|x|\,.
\end{equation}
By applying the differential operators $\hat{\Theta}_{i k}$'s on $\log|x|$, one can see that the final result behaves as  $\frac{1}{|x|^{4}}S_{ijkl}$, where the dimensionless function $S_{ijkl}$ is  transverse and traceless with respect to either $ij$ or $kl$ indices. 

%%%%%%%%%%%%%%%%%%%%%%%%%%%%%%%%%%%%%%%%%%%%%%%%%%%%%%%%%%%%%%%%%%%%%%%%%%%
\subsubsection{Energy-momentum tensor two-point function}
Now we look at the two-point function for energy-momentum tensor. From \cite{Grumiller:2013mxa} we know that $f^{(0)}_{i j}$ is the source for the ordinary energy-momentum tensor, $\tau_{i j}$. So to obtain the two-point function of energy-momentum operator, we have to calculate the variation of $\tau_{i j}$ with respect to $f^{(0)}_{i j}$. The value of $\tau_{i j}$ is given in \cite{Grumiller:2013mxa} (in the following relations, the covariant derivative $D_i$ and $R_{ij}^{\LO}$ belong to the three dimensional boundary space)
	\bea
&&\tau_{ij} = \sigma \big[\tfrac{2}{\ell}\,(E_{ij}^{\TO}+ \tfrac{1}{3} E_{ij}^{\SO}h^{\FO}) -\tfrac4\ell\,E_{ik}^{\SO}\psi^{\FO k}_j
+ \tfrac{1}{\ell}\,h_{ij}^{\LO} E_{kl}^{\SO}\psi_{\FO}^{kl}+ \tfrac{1}{2\ell^3}\,\psi^{\FO}_{ij}\psi_{kl}^{\FO}\psi_{\FO}^{kl}\nn\\ 
&&\quad\,\,
- \tfrac{1}{\ell^3}\,\psi_{kl}^{\FO}\,\big(\psi^{\FO k}_i\psi^{\FO l}_j-\tfrac13\,h^{\LO}_{ij}\psi^{\FO k}_m\psi_{\FO}^{lm}\big)\big] - 4\,D^k B_{ijk}^{\FO} + i\leftrightarrow j\,,
\label{eq:CG17}
\eea
where
\bea\label{eq:CG24}
&&B_{ijk}^{\FO} = \tfrac{1}{2\ell}\,\big(D_j\psi^{\FO}_{ik}-\tfrac12\,h_{ij}^{\LO}\,D^l\psi^{\FO}_{kl}\big) - j \leftrightarrow k \,,\nn\cr &&\cr
&&E^{\TO}_{ij} = -\tfrac{3}{4\ell^2}\,\psi^{\TO}_{ij} -\tfrac{1}{12\ell^{2}}\,h_{ij}^{\LO}\,\psi^{kl}_{\FO} \, \psi_{kl}^{\SO}
-\tfrac{1}{16\ell^{2}}\,\psi^{\FO}_{ij}\,\psi^{\FO}_{kl}\,\psi_{\FO}^{kl} \nonumber \\
&&\qquad - \tfrac{\si}{12}\,\big(R^{\LO}\,\psi_{ij}^{\FO}-h _{ij}^{\LO}\,R_{kl}^{\LO}\,\psi^{kl}_{\FO}
+h _{ij}^{\LO}\,D_{l}\,D_{k}\,\psi^{kl}_{\FO} \nonumber \\
&&\qquad +\tfrac{3}{2}\,D_{k}\,D^{k}\,\psi_{ij}^{\FO}
-3\,D_{k}\,D_{i}\,\psi^{\FO k}_{j} \big) + \tfrac{1}{24\ell^2} \, E_{ij}^{%\textrm{\tiny tr}
\gamma} + i \leftrightarrow j \,,\nn\cr &&\cr
&&E_{ij}^{%\textrm{\tiny tr} 
\gamma} =  h_{\FO}\,(3\,\psi^{\SO}_{ij}+\tfrac12\,h_{ij}^{\LO}\,
\psi_{kl}^{\FO}\,\psi^{kl}_{\FO}-h_{\FO}\,\psi_{ij}^{\FO}) \nonumber \\
&&\quad \,\,\,\, + 5\,h _{\SO}\,\psi_{ij}^{\FO} - \si\ell^2\,(D_{j}\,D_{i}\,h_{\FO} 
-\tfrac{1}{3}\,h _{ij}^{\LO}\,D^{k}\,D_{k}\,h _{\FO})\,.
\label{eq:CG26b}
\eea
As before the first step is to write $f^{(0)}_{i j}$ in a convenient way. Since the only modes including coefficients of $r^{0}$ are the Einstein modes, it will be sufficient to extract such factors from the previously obtained Einstein modes (\ref{4.60}), and use them as our new set of symmetric basis. The result is as follows
	\begin{eqnarray}\label{e.E}
	t^{1,2}_{i j} =(p_{i} \epsilon^{1,2}_{j}+p_{j} \epsilon^{1,2}_{i})
	\,,\qquad
	t^{3,4}_{i j} = M^{1,2}_{i j}
	\,,\qquad
	t^{5}_{i j} = p_{i} p_{j}
	\,,\qquad
	t^{6}_{i j} =-2\eta_{i j}\,.
	\end{eqnarray}
Now, as a generic Einstein solution is written as $f^{E}_{i j}=\sum{B_{I} f^{E{I}}_{i j}}$, it is clear that the $r^{0}$ coefficient in the FG expansion of this mode can be displayed as $f^{(0)}_{i j}=\sum{B_{I} t^{I}_{i j}}$.
So we should find an expression for $\tau_{i j}$ in terms of this basis. From the same considerations as those explained in the case of ghost modes, it turns out that the relevant part of energy-momentum tensor is given by 
\begin{eqnarray}\label{114}
\frac{\delta{\tau_{i j}}}{\delta{f_{(0)}^{k l}}}=6\frac{\delta{\psi^{(3)}_{i j}}}{\delta{f_{(0)}^{k l}}}=6(\frac{\delta{f^{(3)}_{i j}}}{\delta{f_{(0)}^{k l}}}-\frac{1}{3}\eta_{i j}\frac{\delta{f^{(3)}}}{\delta{f_{(0)}^{k l}}})\,. 
\end{eqnarray}
Therefore we should find the contribution of the Einstein modes to $f^{(3)}_{i j}$. That turns out to give
\begin{eqnarray}\label{4.105}
\frac{\delta{f^{(3)}_{i j}}}{\delta{f_{(0)}^{k l}}}=\frac{i|p|^{3}}{3}(t^{3}_{i j}\frac{\delta{B_{3}}}{\delta{f_{(0)}^{k l}}}+t^{4}_{i j}\frac{\delta{B_{4}}}{\delta{f_{(0)}^{k l}}})\,.
	\end{eqnarray}
To compute the final form of the above equation we use the following transverse relations
\begin{eqnarray}
&&t^{K}_{i j} t_{I}^{i j}=-2 {\left|p\right|}^{2} \delta^{K}_{I}\,;\,\,\,K=1,2, \qquad 
t^{K}_{i j} t_{I}^{i j}=2 \delta^{K}_{I}\,;\,\,\, K=3,4,\nn\\
&&t^{5}_{i j} t_{I}^{i j}=p^{4}\delta^{5}_{I}-2p^{2}\delta^{6}_{I}\,,\qquad \quad
t^{6}_{i j} t_{I}^{i j}=-2p^{2}\delta^{5}_{I}+12\delta^{6}_{I}\,,
\end{eqnarray}
where by using (\ref{4.123})	it can be written as 
	\begin{eqnarray}\label{eint}
	&&t^{K}_{i j} f_{(0)}^{i j}=-2 p^{2} B_{K}\,;\,\,\, K=1,2,\qquad
	t^{K}_{i j} f_{(0)}^{i j}=-2 B_{K}\,;\,\,\, K=3,4\,, \nn\\
	&&t^{5}_{i j} f_{(0)}^{i j}=-p^{4}B_{5}+2p^{2}B_{6}\,, \qquad \quad
	t^{6}_{i j} f_{(0)}^{i j}=2p^{2}B_{5}-12B_{6}\,.
	\end{eqnarray}
Therefore equation (\ref{4.105}) becomes 
	\begin{eqnarray}
	\frac{\delta{f^{(3)}_{i j}}}{\delta{f_{(0)}^{k l}}}=-\frac{i|p|^{3}}{6}(t^{3}_{i j}t^{3}_{k l}+t^{4}_{i j}t^{4}_{k l})=-\frac{i}{6|p|}(\Theta_{i k} \Theta_{j l}+\Theta_{i l} \Theta_{j k}-\Theta_{i j} \Theta_{k l})\,,
	\end{eqnarray}
	where we have used (\ref{e.E}) together with (\ref{eint}) and a relation similar to (\ref{330}) for $t^{n}_{i j}$'s in the last equality.
Finally from (\ref{114}) and taking tracelessness of $t^{3}_{i j}$ and $t^{4}_{i j}$ into account, we find the desired two-point function
	\begin{eqnarray}\label{122}
	\big < \tau_{i j} \tau_{k l}\big >=i\frac{\delta{\tau_{i j}}}{\delta{f_{(0)}^{k l}}}=\frac{1}{|p|}(\Theta_{i k} \Theta_{j l}+\Theta_{i l} \Theta_{j k}-\Theta_{i j} \Theta_{k l})\,.
	\end{eqnarray}
By evaluating this two-point function in the configuration space and after regularization we find that
\begin{eqnarray}\label{2pT}
	\langle \tau_{i j} \tau_{k l}\rangle ~=~ i\frac{\delta{\tau_{i j}}}{\delta{f_{(0)}^{k l}}}=4\pi(\hat{\Theta}_{i k} \hat{\Theta}_{j l}+\hat{\Theta}_{i l}\hat{\Theta}_{j k}-\hat{\Theta}_{i j}\hat{\Theta}_{k l})\frac{1}{|x|^{2}}\,.
\end{eqnarray}
This result is consistent with well-known results in \cite{Coriano:2012wp}. It is notable to mention that despite the fact that in the conformal gravity, the transverse and traceless conditions do not hold for the stress tensor in general\cite{Grumiller:2013mxa}, these conditions appear at the second order in metric fluctuations. That can be seen from equations (15) and (27) in \cite{Grumiller:2013mxa}. There, one can see that the trace and divergence of $\tau_{i j}$ begin from the second order in perturbations and therefore do not give rise to any new feature in the result for two-point functions. Therefore in the calculation of the two-point functions, the stress tensor behaves like a transverse-traceless tensor just as before and hence the above result is a consistent one.
%%%%%%%%%%%%%%%%%%%%%%%%%%%%%%%%%%%%%%%%%%%%%%%%%%%%%%%%%%%%%%%%%%%%%%%%%
\subsubsection{Two-point function of PMR with Energy-Momentum Tensor}\label{316}
Now it is straightforward to compute the two-point function $\big< \tau_{i j} P_{k l}\big>$ which can be calculated in two distinct ways either by computing $\frac{\delta{\langle\tau_{i j}\rangle}}{\delta{f_{(1)}^{k l}}}$ or by $\frac{\delta{\langle P_{k l}\rangle}}{\delta{f_{(0)}^{i j}}}$.	
Now it is straightforward to compute the two-point function $\big< \tau_{i j} P_{k l}\big>$ which can be calculated in two distinct ways either by computing $\frac{\delta{\langle\tau_{i j}\rangle}}{\delta{f_{(1)}^{k l}}}$ or by $\frac{\delta{\langle P_{k l}\rangle}}{\delta{f_{(0)}^{i j}}}$.	

To start with the first case, we should determine the relevant part of $\langle\tau_{i j}\rangle$ for this calculation. From (\ref{eq:CG17}) and (\ref{eq:CG26b}) along with bearing in mind our preceding discussion about the relevant order of perturbation and also independence of $f^{(0)}_{i j}$ and $f^{(1)}_{i j}$, we find that
	\begin{eqnarray}
	&&\tau_{i j}|_{relv.}=-2 E^{3}_{i j}-4 D^{k}{B^{1}_{i j k}}+i\leftrightarrow j\,,\nn\\
	&&E^{3}_{i j}|_{relv.}=-\frac{3}{4}\psi^{3}_{i j}+\frac{1}{12}h^{0}_{i j}D_{l}{D_{k}{\psi_{1}^{k l}}}+\frac{1}{8}D_{k}{D^{k}{\psi^{1}_{i j}}}-\frac{1}{4}D_{k}{D_{i}{{\psi^{1k}_{j}}}}\nn \\
	&&\qquad \quad \,\,\,+\frac{1}{24}D_{j}{D_{i}{h_{1}}}-\frac{1}{72}h^{0}_{i j}D^{k}{D_{k}{h_{1}}}+i\leftrightarrow j\,,
	\end{eqnarray}
	while all terms in $B^{1}_{i j k}$ are already relevant. Therefore we explicitly find that
	\begin{eqnarray}\label{4.125}
	\tau_{i j}|_{relv.}\!\!&=&\!\!\big((\frac{3}{2}\psi^{3}_{i j}-\frac{1}{6}h^{0}_{i j}D_{l}{D_{k}{\psi_{1}^{k l}}}-\frac{1}{4}D_{k}{D^{k}{\psi^{1}_{i j}}}+\frac{1}{2}D_{k}{D_{i}{{\psi^{1k}_{j}}}}-\frac{1}{12}D_{j}{D_{i}{h_{1}}}\nn \\
	\!\!&+&\!\!\frac{1}{36}h^{0}_{i j}D^{k}{D_{k}{h_{1}}}+i\leftrightarrow j)-2 D^{k}{D_{j}{\psi^{1}_{i k}}}+ h^{0}_{i j}D^{k}{D^{l}{\psi^{1}_{k l}}}\nn \\
	\!\!&+&\!\! 2 D^{k}{D_{k}{\psi^{1}_{i j}}}-h^{0}_{i k} D^{k}{D^{l}{\psi^{1}_{j l}}}\big)+i\leftrightarrow j\,.
	\end{eqnarray}
We remind that in the above formula, $D_{i}$ indicates covariant derivative in the three dimensional boundary space. Nevertheless, we notice that the only significant part of $D_{i}$, is the ordinary derivative $\partial_{i}$, because the Christoffel symbols are of higher orders in metric fluctuations. 	Regarding (\ref{4.125}), in momentum space we have
	\begin{eqnarray}
	&&\tau_{i j}|_{relv.}=6\psi^{(3)}_{i j}-\frac{4}{3}h^{(0)}_{i j}p_{l}p_{k}\psi_{(1)}^{k l}-3p_{k}p^{k}{\psi^{(1)}_{i j}}+p^{k}p_{i}{\psi^{(1)}_{kj}}+p^{k}p_{j}{\psi^{(1)}_{ki}}\nn \\
	&&\qquad\quad\,+\frac{1}{3}p_{i}p_{j}h_{(1)}-\frac{1}{9}h^{(0)}_{i j}p^{k}p_{k}h_{(1)}+h^{(0)}_{i k} p^{k}p^{l}\psi^{(1)}_{j l}+h^{(0)}_{j k} p^{k}p^{l}\psi^{(1)}_{i l}\,.
	\end{eqnarray}
Now we are ready to compute the desired variation as follows
	\begin{eqnarray}\label{143}
	&&\frac{\delta{\tau_{i j}}}{\delta{f_{(1)}^{k l}}}=6\frac{\delta{\psi^{(3)}_{i j}}}{\delta{f_{(1)}^{k l}}}-\frac{4}{3}\eta_{i j}p^{m}p^{n}\frac{\delta{\psi^{(1)}_{m n}}}{\delta{f_{(1)}^{k l}}}-3p^{2}\frac{\delta{\psi^{(1)}_{i j}}}{\delta{f_{(1)}^{k l}}}+2p^{m}p_{i}\frac{\delta{{\psi^{(1)}_{mj}}}}{\delta{f_{(1)}^{k l}}}\nn \\
	&&\qquad\,\,\,+2p^{m}p_{j}\frac{\delta{{\psi^{(1)}_{mi}}}}{\delta{f_{(1)}^{k l}}}+\frac{1}{3}(p_{i}p_{j}-\frac{1}{3}\eta_{i j}p^{2})\frac{\delta{f_{(1)}}}{\delta{f_{(1)}^{k l}}}\,.
	\end{eqnarray}
To compute the right hand side of the above equation, we should first of all evaluate $\frac{\delta{\psi^{(3)}_{i j}}}{\delta{f_{(1)}^{k l}}}$. Since we are doing variation with respect to $f_{(1)}^{k l}$, we are just dealing with ghost modes in (\ref{4.52}) and therefore we need to calculate their contribution to $\psi^{(3)}_{i j}$. That gives us the following result
\begin{eqnarray}
	&&\frac{\delta{\psi^{(3)}_{i j}}}{\delta{f_{(1)}^{k l}}}=\frac{\delta{f^{(3)}_{i j}}}{\delta{f_{(1)}^{k l}}}-\frac{1}{3}\eta_{i j}\eta^{m n}\frac{\delta{f^{(3)}_{m n}}}{\delta{f_{(1)}^{k l}}}=\frac{p^2}{6}(e^{1}_{i j}\frac{\delta{A_{1}}}{\delta{f_{(1)}^{k l}}}+e^{2}_{i j}\frac{\delta{A_{2}}}{\delta{f_{(1)}^{k l}}})\nn \\
	&&\qquad\,\,\,\,+\frac{p^2}{2}(e^{3}_{i j}\frac{\delta{A_{3}}}{\delta{f_{(1)}^{k l}}}+e^{4}_{i j}\frac{\delta{A_{4}}}{\delta{f_{(1)}^{k l}}})+\frac{p^2}{6}(\frac13\eta_{i j}-\frac{p_{i} p_{j}}{p^{2}})\frac{\delta{A_{5}}}{\delta{f_{(1)}^{k l}}}\,.
	\end{eqnarray}
	At this point we can use  equations (\ref{4.80}) and (\ref{4.81})
	so that we obtain
	\begin{eqnarray}\label{146}
	&&6\frac{\delta{\psi^{(3)}_{i j}}}{\delta{f_{(1)}^{k l}}}=\frac{1}{2p^{2}}(p_{i} p_{k} \Theta_{j l}+p_{i} p_{l} \Theta_{j k}+p_{j} p_{k} \Theta_{i l} +p_{j} p_{l} \Theta_{i k})\nn \\
	&&\qquad\quad\,+\frac{3}{2p^{2}}(\Theta_{i k} \Theta_{j l}+\Theta_{i l} \Theta_{j k}-\Theta_{i j} \Theta_{k l})+\frac{p^2}{18}(\eta_{i j}-3\frac{p_{i} p_{j}}{p^{2}})(\eta_{k l}-3\frac{p_{k} p_{l}}{p^{2}})\,,
	\end{eqnarray}
	which by using (\ref{theta}) can be written as 
	\begin{eqnarray}\label{baba}
	&&6\frac{\delta{\psi^{(3)}_{i j}}}{\delta{f_{(1)}^{k l}}}=\frac{4}{3}p_{k}p_{l}\eta_{i j}-p_{j}p_{l}\eta_{i k}-p_{j}p_{k}\eta_{i l}-p_{i}p_{l}\eta_{j k}-p_{i}p_{k}\eta_{j l}\nn \\
	&&\qquad\quad\,+\frac{3}{2}p^{2}\eta_{i k}\eta_{j l}+\frac{3}{2}p^{2}\eta_{i l}\eta_{j k}+\frac{4}{3}p_{i}p_{j}\eta_{k l}-\frac{13}{9}p^{2}\eta_{i j}\eta_{k l}\,.
	\end{eqnarray}
	We now return to the calculation of other terms in (\ref{143}). As a direct consequence of (\ref{4.122}) along with (\ref{4.80}) and (\ref{4.81}) we will find that
	\begin{equation}\label{148}
	\frac{\delta{{f_{(1)}^{i j}}}}{\delta{f_{(1)}^{k l}}}=\frac{1}{2}\eta^{i}_{k}\eta^{j}_{l}+\frac{1}{2}\eta^{i}_{l}\eta^{j}_{k}-\frac{1}{3}\eta^{i j}\eta_{k l}\,.
	\end{equation}
	Also by using (\ref{148}) one can show that $\frac{\delta\psi^{(1)}_{ij}}{\delta f_{(1)}^{kl}}=
	\frac{\delta f^{(1)}_{ij}}{\delta f_{(1)}^{kl}}$. Substituting (\ref{148}) and (\ref{baba}) in (\ref{143}), gives
	\begin{eqnarray}\label{cortp1}
	\langle\tau_{ij} P_{kl} \rangle = i\frac{\delta{\langle\tau_{i j}\rangle}}{\delta{f_{(1)}^{k l}}}=0\,.
	\end{eqnarray}
This result is consistent with this fact that the correlation function of two operators with different scaling dimensions in a CFT vacuum is zero. Actually in  the calculation of correlation functions, a relevant question is that, in which vacuum one is performing the computations. To answer this question, remind that we considered the linearization of metric around the $AdS_{4}$ vacuum and observed that the on-shell perturbations respect the asymptotically $AdS_{4}$ form. According to asymptotically $AdS_{4}$ form of solutions, the dual field theory is a CFT. Furthermore one may wonder whether for this vacuum spontaneous symmetry breaking has happened or not. To answer this question we should find out what happens to the perturbations if we turn off the sources. It is easy to see from our preceding analysis that the expectation value of both studied operators vanish when we turn off their sources.

As the last step, we are going to compute $\frac{\delta{\langle P_{i j}\rangle}}{\delta{f_{(0)}^{k l}}}$ and of course we expect to get the same result as (\ref{cortp1}). To compute this correlation we use (\ref{70}) and we realize that 
	\begin{eqnarray}\label{153}
	&&\frac{\delta{E^{(2)}_{i j}}}{\delta{f_{(0)}^{k l}}}=-\frac{1}{2}\frac{\delta{\psi^{(2)}_{i j}
	}}{\delta{f_{(0)}^{k l}}}-\frac{1}{2}\frac{\delta{R^{(0)}_{i j}}}{\delta{f_{(0)}^{k l}}}+\frac{1}{6}\eta_{i j}\frac{\delta{R^{(0)}}}{\delta{f_{(0)}^{k l}}}\,,
	\end{eqnarray}
where we have used the fact that, the last term in $E^{(2)}_{i j}$ does not make any contribution, because it is of the second order in $f^{(1)}_{k l}$ and so will  vanish at the end of calculation when we substitute the source $f^{(1)}_{k l}$ with its background value.
	
As before, we start with computation of $\frac{\delta{\psi^{(2)}_{i j}}}{\delta{f_{(0)}^{k l}}}$ while noting that in order to read $\psi^{(2)}_{i j}$, one needs to take just the contribution from the Einstein modes into account, i.e.
	\begin{eqnarray}
	&&\frac{\delta{\psi^{(2)}_{i j}}}{\delta{f_{(0)}^{k l}}}=\frac{\delta{f^{(2)}_{i j}}}{\delta{f_{(0)}^{k l}}}-\frac{1}{3}\eta_{i j}\eta^{m n}
	\frac{\delta{f^{(2)}_{m n}}}{\delta{f_{(0)}^{k l}}}\,.
	\end{eqnarray}
	From (\ref{4.60}) we see that  
	$f^{(2)}_{i j}=-\frac{p^2}{2}B_{3}t^{3}_{i j}-\frac{p^2}{2}B_{4}t^{4}_{i j}+B_{6}t^{5}_{i j}$. 
	Then from (\ref{eint}) we find that 
	\begin{eqnarray}
	&&\frac{\delta{f^{(2)}_{i j}}}{\delta{f_{(0)}^{k l}}}=\frac{p^2}{4}t^{3}_{i j}t^{3}_{k l}+\frac{p^2}{4}t^{4}_{i j}t^{4}_{k l}-\frac{1}{8p^{2}}t^{5}_{i j}(2t^{5}_{k l}+p^{2}t^{6}_{k l})\,.
	\end{eqnarray}
Therefore for $\frac{\delta{\psi^{(2)}_{i j}}}{\delta{f_{0}^{k l}}}$ we find the following result after simplification
\begin{eqnarray}\label{160}
	&&\frac{\delta{\psi^{(2)}_{i j}}}{\delta{f_{(0)}^{k l}}}=\frac{1}{3}p_{k}p_{l}\eta_{i j}-\frac{1}{4}p_{j}p_{l}\eta_{i k}-\frac{1}{4}p_{j}p_{k}\eta_{i l}-\frac{1}{4}p_{i}p_{l}\eta_{j k}-\frac{1}{4}p_{i}p_{k}\eta_{j l}\nn \\
	&&\qquad\,\,\,\,+\frac{1}{2}p_{i}p_{j}\eta_{k l}+\frac{1}{4}p^{2}\eta_{i l}\eta_{j k}+\frac{1}{4}p^{2}\eta_{i k}\eta_{j l}-\frac{1}{3}p^{2}\eta_{i j}\eta_{k l}.
	\end{eqnarray}
To compute $\frac{\delta{R^{(0)}_{i j}}}{\delta{f_{(0)}^{k l}}}$, we exploit the following relation
	\begin{eqnarray}\label{161}
	\frac{\delta{f^{(0)}_{i j}}}{\delta{f_{(0)}^{k l}}}=-\eta_{i m}\eta_{j n}\frac{\delta{f_{(0)}^{m n}}}{\delta{f_{(0)}^{k l}}}=-\frac{1}{2}(\eta_{i k}\eta_{j l}+\eta_{i l}\eta_{j k})\,,
	\end{eqnarray}
	which is a direct consequence of (\ref{4.123}) and (\ref{eint}). 
	That gives us 
	\begin{eqnarray}\label{Riccisat}
	&&\frac{\delta{R^{(0)}_{i j}}}{\delta{f_{(0)}^{k l}}}=\frac{1}{4}\Big(\eta_{j l}p_{k}p_{i}+\eta_{j k}p_{l}p_{i}+\eta_{i k}p_{l}p_{j}+\eta_{i l}p_{k}p_{j}-2\eta_{l k}p_{j}p_{i}-p^{2}(\eta_{i k}\eta_{j l}+\eta_{i l}\eta_{j k})\Big)\,,\nn\\
	&&\frac{\delta{R^{(0)}}}{\delta{f_{(0)}^{k l}}}=\eta^{i j}\frac{\delta{R^{(0)}_{i j}}}{\delta{f_{(0)}^{k l}}}=p_{k}p_{l}-p^{2}\eta_{k l}\,.
	\end{eqnarray}
	Now, based on (\ref{Riccisat}), (\ref{160}) and (\ref{153}) we find that $\frac{\delta{\langle P_{i j}\rangle}}{\delta{\gamma_{0}^{k l}}}$ is vanishing, which is what one has already expected.

For further checks of the results in (\ref{2pP}), (\ref{2pT}) and (\ref{cortp1}) let us show  that these two-point functions satisfy the Ward Identities obtained in \cite{Grumiller:2013mxa} i.e.
\be
h^{(0)}_{ij} \tau^{ij} + \tfrac12\,\psi^{(1)}_{ij} P^{ij} = 0\,, \qquad 
 h^{(0)}_{ij} P^{ij} = 0 \,.
 \ee
 We begin with calculating the variation of the first relation with respect to $h_{(0)}^{k l}$. So we find that
 \be
 \frac{\delta{h^{(0)}_{ij}}}{\delta{h_{(0)}^{k l}}}\tau^{ij}+h^{(0)}_{ij}\frac{\delta{\tau^{ij}}}{\delta{h_{(0)}^{k l}}}+ \tfrac12\,\psi^{(1)}_{ij}\frac{\delta{P^{ij}}}{\delta{h_{(0)}^{k l}}}+\tfrac12\,\frac{\delta{\psi^{(1)}_{ij}}}{\delta{h_{(0)}^{k l}}}P^{ij} = 0 \,.
 \ee
Similar to what we did to find the correlation functions in the previous sections, we firstly calculate the variations in the presence of sources and at the end of the day we turn off the sources. In this way we obtain the equations that two-point functions are supposed to satisfy. 
Now we use the fact that the background value of $h^{(0)}_{i j}$ is  $\eta_{i j}$ and this gives the background value of $\psi^{(1)}_{i j}$ to be zero as well. We also note that $\tau_{i j}$ and $P_{i j}$ vanish when one turns off sources. Therefore the first, third and fourth terms are already seen to be vanished in the absence of sources. For the second term, it is notable that $\frac{\delta{\tau^{ij}}}{\delta{h_{(0)}^{k l}}}$ which gives us the two-point function of Energy-Momentum tensor, has been turned out to be traceless and hence this term also becomes zero.
 
 The next step is to varying the same identity with respect to $h_{(1)}^{k l}$. This gives us
 \be
 \frac{\delta{h^{(0)}_{ij}}}{\delta{h_{(1)}^{k l}}}\tau^{ij}+h^{(0)}_{ij}\frac{\delta{\tau^{ij}}}{\delta{h_{(1)}^{k l}}}+ \tfrac12\,\psi^{(1)}_{ij}\frac{\delta{P^{ij}}}{\delta{h_{(1)}^{k l}}}+\tfrac12\,\frac{\delta{\psi^{(1)}_{ij}}}{\delta{h_{(1)}^{k l}}}P^{ij} = 0 \,.
 \ee
 Again if we turn off the sources, then the first, third and fourth terms give us zero. The second term gives rise to $\eta_{ij}\,\frac{\delta{\tau^{ij}}}{\delta{h_{(1)}^{k l}}}$, which from vanishing of the two-point function of PMR with EM tensor, this turns out to be zero too.
 Therefore the first identity is checked to be consistent with the two-point functions we have found.
 
Now, we turn to the second identity and calculate its variation with respect to $h_{(0)}^{k l}$. That gives us
\be
 \frac{\delta{h^{(0)}_{ij}}}{\delta{h_{(0)}^{k l}}}P^{ij}+h^{(0)}_{ij}\frac{\delta{P^{ij}}}{\delta{h_{(0)}^{k l}}} = 0 \,.
 \ee
 In the absence of sources the first term vanishes because the background value of $P^{ij}$ is zero. The second term also vanishes because it includes the two-point function of PMR with EM tensor.

As another check we calculate the variation of the second identity with respect to $h_{(1)}^{k l}$ 
 \be
 \frac{\delta{h^{(0)}_{ij}}}{\delta{h_{(1)}^{k l}}}P^{ij}+h^{(0)}_{ij}\frac{\delta{P^{ij}}}{\delta{h_{(1)}^{k l}}} = 0\,. 
 \ee
 In this case the second term vanishes as a result of traceless property  of the two-point function of PMR, given by $\frac{\delta{P^{ij}}}{\delta{h_{(1)}^{k l}}}$.
 The first term vanishes again because the background value of $P_{i j}$ is equal to zero.
 
As a final step let us check the following relation found in \cite{Grumiller:2013mxa} 
 \be\label{dddd}
  2 D_{i}\tau^{ij} + 2 D_{i}P^{i}{}_{k}\,h_{(1)}^{k j}+ 2 P^{i}{}_{k}\,D_{i}h_{(1)}^{ kj}= P^{ik} D^{j} h^{(1)}_{i k}\,.
 \ee
 We are going to vary this relation with respect to $h_{(0)}^{k l}$ and then set the sources equal to zero, in this way the two-point functions appear in this relation. We note that the only part of covariant derivative which contributes in this calculation is $\partial_{i}$. One can go further and argue that we  are actually dealing just  with the first term in the left hand side of the above equality. That is because the other three terms are of second order in perturbations and therefore give zero at the end of calculation.
 
 Hence (\ref{dddd}) can be easily checked by doing a Fourier transformation. Then one remains with $2 p_{i}\,\frac{\delta{\tau^{ij}}}{\delta{h_{(0)}^{k l}}}$
 which is equal to zero from (\ref{122}) and the fact that $p_{i}\epsilon^{(I) i}=0$.
  On the other hand, if we do variation with respect to $h_{(1)}^{k l}$ we readily get zero, because the two-point function of PMR with EM tensor is zero. Therefore we showed that our correlation functions satisfy the Weyl and diffeomorphism Ward identities of the boundary theory. 
%%%%%%%%%%%%%%%%%%%%%%%%%%%%%%%%%%%%%%%%%%%%%%%%%%%%%%%%%%%%%%%%%%%%%%%%%%%%%%%
\subsection{Space-like modes}
In this section we are going to make the same analysis for space-like modes. Our ansatz for the metric fluctuations is $h_{\mu \nu}(r,x)=e^{-ip.x} {h}_{\mu \nu}(r)$ where with the choice of $p_{i}=E\delta^{x}_{i}$ for the space-like three-momentum it takes the following form
\be
h_{\mu \nu}(r,x)=e^{-iEx} {h}_{\mu \nu}(r)\,.
\ee
In this case equations in (\ref{rrc}) lead to
\begin{eqnarray}\label{38}
	&&4 r^{2} \textbf{g}''_{r r} +6 r \textbf{g}'_{r r}+(a^{2}-rE^{2}-4)\textbf{g}_{r r}=0\,,\nn\\
	&&4 r^{2} \textbf{g}''_{i r} +6 r \textbf{g}'_{i r} +(a^{2}-rE^2-4) \textbf{g}_{i r}- 4 r \partial_{i}{\textbf{g}_{r r}}=0\,,\nn \\
	&&4 r^{2} \textbf{g}''_{i j} + 6 r \textbf{g}'_{i j} +(a^{2}-rE^2-4) \textbf{g}_{i j}- 4 r \partial_{i}{\textbf{g}_{r j}} - 4 r \partial_{j}{\textbf{g}_{r i}} +8r \eta_{i j} \textbf{g}_{r r}=0\,,\label{40}
\end{eqnarray}
where in writing the above equations we have used constraint relations (\ref{41}) as follows
\bea\label{3838}
&&\textbf{g}_{r x}=-\frac{4ir}{E}\textbf{g}'_{r r}\,,\qquad
\textbf{g}_{t x}=\frac{2i}{E}\textbf{g}_{r t}-\frac{4ir}{E}\textbf{g}'_{r t}\,, \qquad
\textbf{g}_{x x}=\frac{2i}{E}\textbf{g}_{r x}-\frac{4ir}{E}\textbf{g}'_{r x}\,, \nn \\
&&\qquad\quad\,\textbf{g}_{x y}=\frac{2i}{E}\textbf{g}_{r y}-\frac{4ir}{E}\textbf{g}'_{r y}\,, \qquad
\textbf{g}_{y y}=\textbf{g}_{t t}-\textbf{g}_{x x}-4r\textbf{g}_{r r}\,.
\eea
After finding  all of  the ten components from (\ref{38}) and (\ref{3838}), it remains to determine the gauge transformation parameters required to write the solutions in the Gaussian gauge. 
These parameters can be obtained from the following equations
	\begin{eqnarray}
	(r \xi_{r})' =- \frac{r}{2} \textbf{g}_{r r}\,,\quad
(r \xi_{t})' =-r \textbf{g}_{r t}\,,\quad
	(r \xi_{x})' =-r \textbf{g}_{r x} + irE \xi_{r}\,,\quad
	(r \xi_{y})' =-r \textbf{g}_{r y}\,.
	\end{eqnarray}
	Finally we will require to know the metric fluctuation components in the Gaussian gauge 
	\bea
	&&f_{t t}=r(\textbf{g}_{t t}+4 \xi_{r})\,,\qquad
	f_{t x}=r(\textbf{g}_{t x}-iE \xi_{t})\,,\qquad
	f_{t y}=r\textbf{g}_{t y}\,,\nn\\
	&&f_{x x}=r(\textbf{g}_{x x}-2iE \xi_{x}-4 \xi_{r})\,,\quad
	f_{x y}=r(\textbf{g}_{x y}-iE \xi_{y})\,,\quad
	f_{y y}=r(\textbf{g}_{y y}-4 \xi_{r}).
	\eea
	%%%%%%%%%%%%%%%%%%%%%%%%%%%%%%%%%%%%%%%%%%%%%%%%%%%%%%%%%%%%%%%%%%%%
\subsubsection{Ghost modes}
Let us start to write the result for metric fluctuations in the Gaussian gauge for ghost modes
		\bea
	&&f_{t t}(r)=-c_{2}\frac{4}{E^{4}} e^{-E\sqrt{r}}(2+E\sqrt{r})-c_{1}\frac{1}{E}\sqrt{r}e^{-E\sqrt{r}}+4c_{5}\,,\nn\\
	&&f_{t x}(r)=c_{4}\frac{4i}{E^{3}} e^{-E\sqrt{r}}-iEc_{3}\,,\quad
	f_{t y}(r)=-c_{8}\frac{1}{E}\sqrt{r}e^{-E\sqrt{r}}\,,\nn\\
	&&f_{x x}(r)=-c_{2}\frac{8}{E^{4}} e^{-E\sqrt{r}}+2(E^{2}r-2)c_{5}-2iEc_{7}\,,\nn\\
	&&f_{y y}(r)=c_{2}\frac{4}{E^{4}} e^{-E\sqrt{r}}(2+E\sqrt{r})-c_{1}\frac{1}{E}\sqrt{r}e^{-E\sqrt{r}}-4c_{5}\,,\nn\\
	&&f_{x y}(r)=c_{6}\frac{4i}{E^{3}} e^{-E\sqrt{r}}-iEc_{9}\,.
	\eea
	These solutions have been figured out by taking into account the regularity condition for space-like modes inside the bulk, i.e. we demand that solutions stay finite as the holographic coordinate $r$ goes to infinity and do not diverge \cite{Son:2002sd}.
We choose a specific combination of the above constants $c_{n}$'s, so that  no coefficients of $r^{0}$ in the $r$-expansion of the ghost modes appear
\bea
&&f^{G(1)}_{{i j}} =(1-e^{-E \sqrt{r}}) \mathcal{M}_1\,,\qquad
f^{G(2)}_{{i j}} =(1-e^{-E \sqrt{r}}) \mathcal{M}_4\,,\nn\\
&&
f^{G(3)}_{{i j}} =\sqrt{r} e^{-E \sqrt{r}} \mathcal{M}_2\,,\qquad
f^{G(4)}_{{i j}} =\sqrt{r} e^{-E \sqrt{r}} \mathcal{M}_7\,,\nn\\
&&
f^{G(5)}_{{i j}} =\big(\frac{2}{E} (e^{-E \sqrt{r}}-1)+\sqrt{r} e^{-E \sqrt{r}}\big) \mathcal{M}_6 - \big(\frac{2}{E} (e^{-E \sqrt{r}}-1)-E r\big) \mathcal{M}_8\,,
\eea
where $\mathcal{M}_7=\frac12(\mathcal{I}-\mathcal{M}_3+\mathcal{M}_5)$ and
$\mathcal{M}_8=\frac12(\mathcal{I}+\mathcal{M}_3-\mathcal{M}_5)$.
Therefore we can write the solution for a general space-like three-momentum $p_{i}$ as follows
	\begin{eqnarray}\label{f12}
	&&f^{G(1,2)}_{i j} =(1-e^{-{\left|p\right|} \sqrt{r}}) \frac{1}{\left|p\right|}(p_{i} \epsilon^{1,2}_{j}+p_{j} \epsilon^{1,2}_{i})\,,\qquad
	\label{f34}
	f^{G(3,4)}_{i j} =\sqrt{r} e^{-{\left|p\right|} \sqrt{r}} M^{1,2}_{i j}\,,\nn\\
	&&\label{f5}
	f^{G(5)}_{i j} =\frac{2}{{\left|p\right|}} (e^{-{\left|p\right|} \sqrt{r}}-1) (\eta_{i j}-2\frac{p_{i} p_{j}}{p^2})
	+\sqrt{r} e^{-{\left|p\right|} \sqrt{r}} (\eta_{i j}-\frac{p_{i} p_{j}}{p^2})
  +{\left|p\right|} r \frac{p_{i} p_{j}}{p^2}\,,
  \end{eqnarray}
	where in the above relations, $\epsilon^{1}_{i}$ and $\epsilon^{2}_{i}$ are two transverse polarizations of the spin-2 field, from which two symmetric and traceless tensors $M^{1,2}_{i j}$ are constructed as 
	\begin{eqnarray}
	M^{1}_{i j}=\epsilon^{1}_{i} \epsilon^{1}_{j}+\epsilon^{2}_{i} \epsilon^{2}_{j}\,,\qquad 
	M^{2}_{i j}=\epsilon^{1}_{i} \epsilon^{2}_{j}+\epsilon^{2}_{i} \epsilon^{1}_{j}\,.
	\end{eqnarray}
 Having found these solutions, it is possible  to repeat the analysis of the preceding subsections to calculate the desired two-point functions for the space-like modes. 
%%%%%%%%%%%%%%%%%%%%%%%%%%%%%%%%%%%%%%%%%%%%%%%%%%%%%%%%%%%%%%%%%%%%%%
\subsubsection{Einstein modes}
The results for metric fluctuations in Gaussian gauge for Einstein modes are given by
	\bea
&&f_{t t}(r)=-c_{1}\frac{1}{E^{3}} e^{-E\sqrt{r}}(1+E\sqrt{r})+4c_{2}\,,\nn \\
&&f_{t x}(r)=-iEc_{3}\,,\quad
f_{t y}(r)=c_{6}\frac{3i}{2E^{2}}e^{-E\sqrt{r}}(1+E\sqrt{r})\,,\nn\\
&&f_{x x}(r)=2(-2+E^{2}r)c_{2}-2iEc_{5}\,,\quad f_{x y}(r)=-iEc_{4}\,,\nn\\
&&f_{y y}(r)=-c_{1}\frac{1}{E^{3}} e^{-E\sqrt{r}}(1+E\sqrt{r})-4c_{2}\,.
\eea
As we noted before these solutions have been figured out by taking into account the regularity condition for space-like modes inside the bulk. It is possible to choose a specific combination of the above constants $c_{n}$'s, so that the following results will appear
\bea
&&f^{E(1)}_{i j} =E \mathcal{M}_1\,,\qquad
f^{E(2)}_{i j} =E \mathcal{M}_4\,,\nn\\
&&
f^{E(3)}_{i j} =(1+E\sqrt{r})e^{-E \sqrt{r}} \mathcal{M}_2\,,\quad
f^{E(4)}_{i j} =(1+E\sqrt{r})e^{-E \sqrt{r}} \mathcal{M}_7\,,\nn\\
&&
f^{E(5)}_{i j} = E^2 \mathcal{M}_8\,,\qquad
f^{E(6)}_{i j} = \mathcal{M}_6 + (1-\frac{E^2 r}{2}) \mathcal{M}_8\,.
\eea
Therefore we can write the solutions for a general space-like three-momentum $p_{i}$ as follows
	\begin{eqnarray}\label{E12}
&&	f^{E(1,2)}_{i j} =p_{i} \epsilon^{1,2}_{j}+p_{j} \epsilon^{1,2}_{i}
	\,,\quad\label{E34}
	f^{E(3,4)}_{i j} =(1+|p|\sqrt{r})\sqrt{r} e^{-{\left|p\right|} \sqrt{r}} M^{1,2}_{i j}
	\,,\nn\\ \label{E5}
&&	f^{E(5)}_{i j} =p_{i}p_{j}
  \,,\qquad\qquad\qquad\label{E6}
	f^{E(6)}_{i j} =\eta_{i j}-\frac{r}{2}p_{i}p_{j}\,.
  \end{eqnarray}
%%%%%%%%%%%%%%%%%%%%%%%%%%%%%%%%%%%%%%%%%%%%%%%%%%%%%%%%%%%%%%%%%%%%%%%
 \subsubsection{Two-point function of PMR}
 We have already obtained the five-fold basis from (\ref{f12}) as follows
  \begin{eqnarray}
	e^{1,2}_{i j} =(p_{i} \epsilon^{1,2}_{j}+p_{j} \epsilon^{1,2}_{i})\,,\qquad
	e^{3,4}_{i j} = M^{1,2}_{i j}\,,\qquad
	e^{5}_{i j} =(\frac{3p_{i} p_{j}}{p^2}-\eta_{i j})\,.
	\end{eqnarray}     
Then we should find an expression for $P_{i j}$, in terms of this basis. We find that
\be
	f^{G(2)}_{i j}=-\frac {\left|p\right|}{2} A_{1} e^{1}_{i j}-\frac{\left|p\right|}{2} A_{2} e^{2}_{i j}- {\left|p\right|} A_{3} e^{3}_{i j}-{\left|p\right|} A_{4} e^{4}_{i j}\,.
\ee
The transverse relations can be written as
	\begin{eqnarray}
	&& e^{1}_{i j} e_{I}^{i j}=-2p^{2}\delta^{1}_{I}\,,\quad
	e^{2}_{i j} e_{I}^{i j}=2p^{2}\delta^{2}_{I}\,,\quad
	e^{3}_{i j} e_{I}^{i j}=-2\delta^{3}_{I}\,,\nn\\
	&& e^{4}_{i j} e_{I}^{i j}=2\delta^{4}_{I}\,, \quad
	e^{5}_{i j} e_{I}^{i j}=6\delta^{5}_{I}.
	\end{eqnarray}
By doing the same analysis as the time-like case, we get 
	\begin{eqnarray}\label{44}
	\frac{\delta{f^{(2)}_{i j}}}{\delta{f_{(1)}^{k l}}}=\frac{1}{4{\left|p\right|}}(e^{1}_{i j} e^{1}_{k l}-e^{2}_{i j} e^{2}_{k l})+\frac{\left|p\right|}{2} (e^{3}_{i j} e^{3}_{k l}-e^{4}_{i j} e^{4}_{k l})\,.
	\end{eqnarray}	
Now if we write the $e^{I}_{i j}$'s in the above formula in terms of polarization vectors and use the relation $\epsilon^{1}_{i} \epsilon^{1}_{j}-\epsilon^{2}_{i} \epsilon^{2}_{j}=\frac{p_{i} p_{j}}{p^2}-\eta_{i j}$ then we will find that
	\begin{eqnarray}
	&&e^{1}_{i j} e^{1}_{k l}-e^{2}_{i j} e^{2}_{k l}=-\frac{1}{p^{2}}(p_{i} p_{k} \Theta_{j l}+p_{i} p_{l} \Theta_{j k}+p_{j} p_{k} \Theta_{i l} +p_{j} p_{l} \Theta_{i k})\,,\nn \\
	&&e^{3}_{i j} e^{3}_{k l}-e^{4}_{i j} e^{4}_{k l}=\frac{1}{p^{4}}(\Theta_{i j} \Theta_{k l}-\Theta_{i k} \Theta_{j l}-\Theta_{i l} \Theta_{j k})\,.
	\end{eqnarray}
	Therefore one can rewrite (\ref{44}) in the following form
	\begin{eqnarray}
	\frac{\delta{f^{(2)}_{i j}}}{\delta{f_{(1)}^{k l}}}\!\!&=&\!\!-\frac{1}{4 {\left|p\right|}^3}(p_{i} p_{k} \Theta_{j l}+p_{i} p_{l} \Theta_{j k}+p_{j} p_{k} \Theta_{i l} +p_{j} p_{l} \Theta_{i k})\nn\\
	\!\!&-&\!\!\frac{1}{2 {\left|p\right|}^3} (\Theta_{i k} \Theta_{j l}+\Theta_{i l} \Theta_{j k}-\Theta_{i j} \Theta_{k l})\,,
	\end{eqnarray}
So the two-point function for PMR is given by
	\begin{eqnarray}
	\langle P_{i j}P_{k l} \rangle \!\!&=&\!\! i\frac{\delta{\left\langle P_{i j}\right\rangle}}{\delta{f_{(1)}^{k l}}}=\frac{i}{2 {\left|p\right|}^3}(p_{i} p_{k} \Theta_{j l}+p_{i} p_{l} \Theta_{j k}+p_{j} p_{k} \Theta_{i l} +p_{j} p_{l} \Theta_{i k})\nn\\
	\!\!&+&\!\!\frac{i}{{\left|p\right|}^3} (\Theta_{i k} \Theta_{j l}+\Theta_{i l} \Theta_{j k}-\Theta_{i j} \Theta_{k l})\,,
	\end{eqnarray} 
where up to an $i$ factor is the same as the result for the time-like case in (\ref{2pP}).
%%%%%%%%%%%%%%%%%%%%%%%%%%%%%%%%%%%%%%%%%%%%%%%%%%%%%%%%%%%%%%%%%%%%%%%%%%%%%
\subsubsection{Two-point function of energy-momentum tensor}
 In this case we obtain our six-fold basis from (\ref{E12})  as 
  \begin{eqnarray}
	t^{1,2}_{i j} =(p_{i}\epsilon^{1,2}_{j}+p_{j}\epsilon^{1,2}_{i})\,,\qquad
	t^{3,4}_{i j} = M^{1,2}_{i j}\,,\qquad
	t^{5}_{i j} =p_{i}p_{j}\,,\qquad t^{6}_{i j} =\eta_{i j}\,.
	\end{eqnarray}
	Therefore $f^{(0)}_{i j}$ evidently is written as $\sum{B_{I}t^{I}_{i j}}$ and $f^{(3)}_{i j}$ turns out to be 
	\begin{eqnarray}
	f^{(3)}_{i j}=\frac{|p|^{3}}{3}(B_{3}t^{3}_{i j}+B_{4}t^{4}_{i j})\,.
	\end{eqnarray}
	In this case the transverse relations take the following forms 
	\begin{eqnarray}\label{69}
	&&t^{1}_{i j} t_{I}^{i j}=-2 p^{2} \delta^{1}_{I}\,,\qquad
	t^{2}_{i j} t_{I}^{i j}=2 p^{2} \delta^{2}_{I}\,,\qquad
	t^{3}_{i j} t_{I}^{i j}=-2\delta^{3}_{I}\,,	\nn\\
	&&t^{4}_{i j} t_{I}^{i j}=2\delta^{4}_{I}\,,\qquad
	t^{5}_{i j} t_{I}^{i j}=p^{4}\delta^{5}_{I}+p^{2}\delta^{6}_{I}\,,\qquad 
	t^{6}_{i j}t_{I}^{i j}=p^{2}\delta^{5}_{I}+3\delta^{6}_{I}\,.
	\end{eqnarray}
	from which it follows that 
	\begin{eqnarray}
&&\frac{\delta{B_{1}}}{\delta{f_{(0)}^{i j}}}=\frac{1}{2p^2}t^{1}_{i j}\,,\qquad
\frac{\delta{B_{2}}}{\delta{f_{(0)}^{i j}}}=-\frac{1}{2p^2}t^{2}_{i j} \,,\qquad
\frac{\delta{B_{3}}}{\delta{f_{(0)}^{i j}}}=\frac{1}{2}t^{3}_{i j}\,,\qquad
\frac{\delta{B_{4}}}{\delta{f_{(0)}^{i j}}}=-\frac{1}{2}t^{4}_{i j} \,,\nn \\
&&\frac{\delta{B_{5}}}{\delta{f_{(0)}^{i j}}}=-\frac{1}{2p^4}(3t^{5}_{i j}-p^{2}t^{6}_{i j})\,,\qquad
\frac{\delta{B_{6}}}{\delta{f_{(0)}^{i j}}}=\frac{1}{2p^2}(t^{5}_{i j}-p^{2}t^{6}_{i j})\,.
	\end{eqnarray}
	Now by starting from equation (\ref{114}) we have
	\begin{eqnarray}\label{T}
	\langle \tau_{ij} \tau_{kl} \rangle \!\! &=& \!\!i\frac{\delta{\left\langle \tau_{i j}\right\rangle}}{\delta{f_{(0)}^{k l}}}=2i|p|^{3}(\frac{\delta{B_{3}}}{\delta{f_{(0)}^{k l}}}t^{3}_{i j}+\frac{\delta{B_{4}}}{\delta{f_{(0)}^{k l}}}t^{4}_{i j})\nn\\
	\!\!&=&\!\!-\frac{i}{|p|}(\Theta_{i k} \Theta_{j l}+\Theta_{i l} \Theta_{j k}-\Theta_{i j} \Theta_{k l})\,, 
	\end{eqnarray}
which is a result very similar to time-like case in equation (\ref{122}).
%%%%%%%%%%%%%%%%%%%%%%%%%%%%%%%%%%%%%%%%%%%%%%%%%%%%%%%%%%%%%%%%%%%%%%%%%%%%%%
\subsubsection{Two-point function of PMR with energy-momentum tensor}
We can also find the two-point function $\langle \tau_{i j} P_{k l}\rangle$ by calculating $\frac{\delta{\left\langle \tau_{i j}\right\rangle}}{\delta{\gamma_{1}^{k l}}}$. We  again use the relation (\ref{143}).
First of all one should evaluate $\frac{\delta{f^{(3)}_{i j}}}{\delta{f_{(1)}^{k l}}}$. The result turns out to be 
	\begin{eqnarray}
	&&\frac{\delta{f^{(3)}_{i j}}}{\delta{f_{(1)}^{k l}}}=-\frac{1}{12}(e^{1}_{i j} e^{1}_{k l}-e^{2}_{i j} e^{2}_{k l})-\frac{p^{2}}{4}(e^{3}_{i j} e^{3}_{k l}-e^{4}_{i j} e^{4}_{k l})+\frac{p^{2}}{36}(\eta_{i j}+\frac{p_{i}p_{j}}{p^{2}})e^{5}_{k l}\,,
	\end{eqnarray}
	therefore we find that
	\begin{eqnarray}
	&&6\frac{\delta{\psi^{3}_{i j}}}{\delta{f_{1}^{k l}}}=\frac{1}{2p^{2}}(p_{i} p_{k} \Theta_{j l}+p_{i} p_{l} \Theta_{j k}+p_{j} p_{k} \Theta_{i l} +p_{j} p_{l} \Theta_{i k})\nn \\
	&&+\frac{3}{2p^{2}}(\Theta_{i k} \Theta_{j l}+\Theta_{i l} \Theta_{j k}-\Theta_{i j} \Theta_{k l})
	+\frac{p^2}{18}(\eta_{i j}-3\frac{p_{i} p_{j}}{p^{2}})(\eta_{k l}-3\frac{p_{k} p_{l}}{p^{2}})
	\,,
	\end{eqnarray}
which is the same as (\ref{146}).
For other terms in (\ref{143}) one can easily verify that (\ref{148}) is still valid. Therefore it is concluded that the two-point correlation function of PMR and Energy-Momentum tensor is again vanishing
\be
\langle \tau_{i j} P_{k l}\rangle ~=~ 0\,.
\ee
One may check this result by using an alternative way. The value of $f^{(2)}_{i j}$ is given by
\bea
f^{(2)}_{i j}=-\frac{p^{2}}{2}(B_{3}t^{3}_{i j}+B_{4}t^{4}_{i j})-\frac{1}{2}B_{6}p_{i}p_{j}\,.
\eea
Therefore by using the equation (\ref{69}) we find that
\begin{eqnarray}
	\frac{\delta{\psi^{(2)}_{i j}}}{\delta{f_{(0)}^{k l}}}=\frac{1}{4p^{2}}(\Theta_{i k} \Theta_{j l}+\Theta_{i l} \Theta_{j k}-\Theta_{i j} \Theta_{k l})-\frac{1}{4p^{2}}(t^{5}_{i j}-\frac{p^{2}}{3}t^{6}_{i j})(t^{5}_{k l}-p^{2}t^{6}_{k l})\,, 
	\end{eqnarray}
	which can be checked to be the same as (\ref{160}). Moreover it is verified that the equation (\ref{161}) is also valid for this case. Then by considering equation (\ref{153}) we see that the correlation function of PMR with energy-momentum tensor is equal to zero which is a consistent result.
	%%%%%%%%%%%%%%%%%%%%%%%%%%%%%%%%%%%%%%%%%%%%%%%%%%%%%%%%%%%%%%
\subsection{Light-like modes}
	To complete this note, it remains for us to repeat this analysis for the case of light-like modes. Since there is nothing new in the main stream of the calculations, we present it briefly by directly writing the Einstein and ghost modes which we have obtained as solutions to the linearized Bach equation in the Gaussian gauge.
	The ghost solutions are 
	\bea
	f_{t t}\!\!\!&=&\!\!\!-2c_{5}r^{\frac12}+(c_{6}+2E^{2}c_{11})r+\frac{16}{3}i (E c_{3}+i c_{1})r^{\frac{3}{2}}-\frac{1}{3}(c_{2}+iE c_{4})r^{2}
	+\frac{E^{2}}{90}c_{2}r^{3}+4c_{11}\,,\nn\\
	f_{t x}\!\!\!&=&\!\!\!-(\frac{8ic_{3}}{E} \!+\! 2c_{5})r^{\frac12} \!+\! (\frac{2ic_{4}}{E} \!+\! c_{6} \!+\! 2E^{2}c_{11})r\!+\! \frac{16}{3}(i E c_{3} \!-\! c_{1})r^{\frac{3}{2}}
	\!-\! \frac{iE c_{4}}{3} r^{2} \!+\! \frac{E^{2}c_{2}}{90}r^{3} \!-\!  iEc_{12},\nn\\	
	f_{t y}\!\!\!&=&\!\!\!(\frac{8i}{E}c_{7}-2c_{9})r^{\frac12}+(c_{10}-\frac{2i}{E}c_{8})r+\frac{8}{3}iE c_{7}r^{\frac{3}{2}}\!-\!
	\frac{1}{6}iEc_{8}r^{2}\,,\nn\\
	f_{x y}\!\!\!&=&\!\!\!-2c_{9}r^{\frac12}+c_{10}r+\frac{8}{3}iE c_{7}r^{\frac{3}{2}}-\frac{1}{6}iEc_{8}r^{2}\,,\nn\\	
	f_{x x}\!\!\!&=&\!\!\!\frac{16}{E^{2}}(c_{1}\!-\!iEc_{3}\!-\!\frac{E^{2}c_{5}}{8})r^{\frac12}\!+\!(\frac{4ic_{4}}{E}\!+\!c_{6}\!+\!2E^{2}c_{11})r\!+\!\frac{16}{3}i(Ec_{3}\!+\!ic_{1})r^{\frac{3}{2}}\!-\!\frac{1}{3}(c_{2}\!+\!iE c_{4})r^{2}\nn\\
	\!\!\!&+&\!\!\!\frac{E^{2}}{90}c_{2}r^{3}\!-\!4c_{11}\!-\!2iEc_{12}\,,\nn\\
	f_{y y}\!\!\!&=&\!\!\!\frac{16}{E^{2}}(-c_{1}+iEc_{3})r^{\frac12}-\frac{4i}{E}c_{4}r+\frac{16}{3}c_{1}r^{\frac{3}{2}}-\frac{1}{3}c_{2}r^{2}-4c_{11}\,.
	\eea
	Now if we impose the regularity boundary condition in the bulk we can show that $c_{1}$, $c_{2}$, $c_{3}$, $c_{4}$, $c_{7}$ and $c_{8}$ should be equal to zero. Therefore we remain just with the following modes 
	\begin{eqnarray}
		f^{G(1)}_{i j} =\sqrt{r}
		\begin{pmatrix}
		1 & 1 & 0 \\
		1 & 1 & 0 \\
		0 & 0 & 0 \\
		\end{pmatrix}\,,&f^{G(2)}_{i j} =\sqrt{r}
		\begin{pmatrix}
		0 & 0 & 1 \\
		0 & 0 & 1 \\
		1 & 1 & 0 \\
		\end{pmatrix}\,,\nn\\
		f^{G(3)}_{i j} =r
		\begin{pmatrix}
		1 & 1 & 0 \\
		1 & 1 & 0 \\
		0 & 0 & 0 \\
		\end{pmatrix}\,,&f^{G(4)}_{i j} =r
		\begin{pmatrix}
		0 & 0 & 1 \\
		0 & 0 & 1 \\
		1 & 1 & 0 \\
		\end{pmatrix}\,.
		\end{eqnarray}
	So if we consider $p_{i}=(E,E,0)$, $\epsilon^{1}_{i}=(0,0,1)$ and $\epsilon^{2}_{i}=(1,1,0)$, then we can write the solution for a general light-like three-momentum $p_{i}$ as
		\begin{eqnarray}\label{lE1}
	&&	f^{G(1)}_{i j} =r^{\frac12} \epsilon^{2}_{i}\epsilon^{2}_{j}\,,\qquad
		\label{lE2}
		f^{G(2)}_{i j} =r^{\frac12} (\epsilon^{1}_{i}\epsilon^{2}_{j}+\epsilon^{2}_{i}\epsilon^{1}_{j})\,,\quad \nn\\
		\label{lE3} 
	&&	f^{G(3)}_{i j} =r\epsilon^{2}_{i}\epsilon^{2}_{j}\,,\qquad\,\,\,
		\label{lE4}
		f^{G(4)}_{i j} =r(\epsilon^{1}_{i}\epsilon^{2}_{j}+\epsilon^{2}_{i}\epsilon^{1}_{j}).
		  \end{eqnarray}
	From the form of solutions it is easy to see that the light-like  mode $f^{(2)}_{i j}$, does not depend on $f^{(1)}_{i j}$. Hence, we have $\frac{\delta{f^{(2)}_{i j}}}{\delta{f_{(1)}^{i j}}}=0$. On account of this and also from (\ref{70}) one finds that the light-like modes do not contribute to the two-point function of PMR operator.

	Now let's look at the two-point function of PMR with EM tensor. In this case, we are dealing with (\ref{143}). Notably, one recognizes that the first term in (\ref{143}) does not exist for light-like modes as a result of regularity condition in the bulk. For other terms, one should know the quantity $\frac{\delta{f^{(1)}_{i j}}}{\delta{f_{(1)}^{k l}}}$, which turns out to be 
	\bea
\frac{\delta{f^{(1)}_{i j}}}{\delta{f_{(1)}^{k l}}}=e^{1}_{i j}e^{1}_{k l}+\frac{1}{2}e^{2}_{i j}e^{2}_{k l}\,;\qquad
e^{1}_{i j}=\epsilon^{2}_{i}\epsilon^{2}_{j}\,,\qquad e^{2}_{i j}=\epsilon^{1}_{i}\epsilon^{2}_{j}+\epsilon^{2}_{i}\epsilon^{1}_{j}\,.
	\eea
	From the above expression  and taking into account that $p^{2}=0$ and $p.\epsilon^{i}=0$ for $i=1,2$, it can be verified that equation (\ref{143}) is simply vanishing. Therefore the light-like modes do not enter into the calculation of two-point function of PMR with energy-momentum tensor as well.

Finally For Einstein solutions, after imposing regularity conditions, we obtain 
\bea
&&f_{t t}=c_{6}+6iEc_{4}r\,,\qquad
f_{t x}=c_{6}+6i(\frac{1}{E}+Er)c_{4}\,,\qquad
f_{t y}=c_{1}+3iEc_{8}r\,,\nn\\
&&f_{x x}=c_{6}+6i(\frac{2}{E}+Er)c_{4}\,,\qquad
f_{x y}=c_{1}+3iEc_{8}r\,,\qquad
f_{y y}=-\frac{12i}{E}c_{4}\,.
\eea
Consequently one finds the following modes 
\bea\label{lab}
&&f^{(1)}_{i j}=\epsilon^{2}_{i}\epsilon^{2}_{j}\,,\qquad \qquad
f^{(2)}_{i j}=\epsilon^{1}_{i}\epsilon^{2}_{j}+\epsilon^{2}_{i}\epsilon^{1}_{j}\,,\nn\\
&&f^{(3)}_{i j}=-3\epsilon^{1}_{i}\epsilon^{1}_{j}+\epsilon^{2}_{i}\epsilon^{2}_{j}+\eta_{i j}+p_{i}p_{j}\,r \,,\quad
f^{(4)}_{i j}=(\epsilon^{1}_{i}\epsilon^{2}_{j}+\epsilon^{2}_{i}\epsilon^{1}_{j})r\,.
\eea
It is clear that in this case by imposing the regularity boundary condition and hence killing all the coefficients of $r^{\frac{3}{2}}$, we find that the two-point correlation function of energy-momentum tensor vanishes, which is verified from equation (\ref{114}).

As the final task one should check that the two-point function of PMR with EM tensor again turns out to be zero. This can be done by using equation (\ref{153}). Let us first compute $\frac{\delta{\psi^{(2)}_{i j}}}{\delta{f_{(0)}^{k l}}}$. Our traceless basis is as follow
\bea\label{o}
t^{1}_{i j}=\epsilon^{2}_{i}\epsilon^{2}_{j}\,,\qquad
t^{2}_{i j}=\epsilon^{1}_{i}\epsilon^{2}_{j}+\epsilon^{2}_{i}\epsilon^{1}_{j}\,,\qquad
t^{3}_{i j}=\epsilon^{2}_{i}\epsilon^{2}_{j}+\eta_{i j}-3\epsilon^{1}_{i}\epsilon^{1}_{j}\,.
\eea
From (\ref{lab}) we find that $f^{(2)}_{i j}=B_{3}\,p_{i}p_{j}+B_{4}t^{2}_{i j}$ so we conclude that
\be
\frac{\delta{f^{(2)}_{i j}}}{\delta{f_{(0)}^{k l}}}=p_{i}p_{j}\frac{\delta{B_{3}}}{\delta{f_{(0)}^{k l}}}+t^{2}_{i j}\frac{\delta{B_{4}}}{\delta{f_{(0)}^{k l}}}\,.
\ee
Now by using equation (\ref{o}) we see that $t^{3}_{i j}t_{3}^{i j}=6$, 
while the other contractions are all equal to zero. Therefore $t^{3}_{i j}t_{I}^{i j}=6\delta^{3}_{I}$ and hence from $f_{(0)}^{i j}=-(B_{1}t_{1}^{i j}+B_{2}t_{2}^{i j}+B_{3}t_{3}^{i j})$, we find that
\be
\frac{\delta{B_{3}}}{\delta{f_{(0)}^{i j}}}=-\frac{1}{6}t^{3}_{i j}\,,\qquad 
\frac{\delta{B_{4}}}{\delta{f_{(0)}^{i j}}}=0\,.
\ee
Then we obtain 
\be
\frac{\delta{\psi^{(2)}_{i j}}}{\delta{f_{(0)}^{k l}}}=\frac{\delta{f^{(2)}_{i j}}}{\delta{f_{(0)}^{k l}}}=-\frac{1}{6}t^{3}_{k l}p_{i}p_{j}\,.
\ee
Now let us to compute $\frac{\delta{{f_{(0)}}_{i j}}}{\delta{f_{(0)}^{k l}}}$. This leads to 
\be\label{ol}
\frac{\delta{f^{(0)}_{i j}}}{\delta{f_{(0)}^{k l}}}=t^{3}_{i j}\frac{\delta{B_{3}}}{\delta{f_{(0)}^{k l}}}=-\frac{1}{6}t^{3}_{i j}t_{3}^{k l}.
\ee
Also we have 
\be
\frac{\delta{R^{(0)}_{i j}}}{\delta{f_{(0)}^{k l}}}=\frac{1}{2}(-p^{n}p_{i}\,\frac{\delta{f^{(0)}_{n j}}}{\delta{f_{(0)}^{k l}}}-p^{n}p_{j}\,\frac{\delta{f^{(0)}_{n i}}}{\delta{f_{(0)}^{k l}}}+p_{i}p_{j}\eta^{a n}\frac{\delta{f^{(0)}_{a n}}}{\delta{f_{(0)}^{k l}}})\,.
\ee
The last term in the above equation does not contribute since the trace of (\ref{ol}) over $ij$ indices gives zero and so the result is 
\be
\frac{\delta{R^{(0)}_{i j}}}{\delta{f_{(0)}^{k l}}}=\frac{1}{12}p^{n}p_{i}t^{3}_{n j}t^{3}_{k l}+\frac{1}{12}p^{n}p_{j}t^{3}_{n i}t^{3}_{k l}=\frac{1}{6}p_{i}p_{j}t^{3}_{k l}\,,
\ee
from which we find that ($p^{2}=0$) 
\be
\frac{\delta{R^{(0)}}}{\delta{f_{(0)}^{k l}}}=0\,.
\ee
All in all, from (\ref{153}) and the above results, it turns out that correlation function of PMR with energy-momentum tensor is vanishing as expected. That brings us to the end of this calculation. The conclusion of this section is that the light-like modes do not enter in the calculation of the two-point functions.

%%%%%%%%%%%%%%%%%%%%%%%%%%%%%%%%%%%%%%%%%%%%%%%%%%%%%%
\section{Summary and Conclusions}

By using the asymptotic analysis, authors of \cite{Grumiller:2013mxa} had shown that by considering some specific boundary conditions for conformal gravity, one can consistently identify the first two coefficients of the Fefferman-Graham expansion of the metric fluctuation (\ref{exp}), i.e. $h^{(0)}_{i j}$ and $h^{(1)}_{i j}$, with two independent sources corresponding to two operators in the QFT side. These operators are the energy-momentum tensor, $\tau_{i j}$, and the partially massless response, $P_{i j}$ respectively. Then the one-point functions have been computed. These one-point functions turn out as some cumbersome functions of $h^{(3)}_{i j}$ and $h^{(2)}_{i j}$ as expected \cite{Grumiller:2013mxa}.

Here we went one step further and solved the linearized equation of motion of conformal gravity inside the bulk. That is crucial in order to find the functional dependence of different coefficients of FG expansion in terms of each other and to be able to vary
the one-point functions with respect to sources and computing the two-point functions.

This computation has been performed in the momentum space of the boundary coordinates; hence we were just dealing with ODEs in the holographic coordinate $r$. We perform this analysis for three different cases of time-like, space-like and light-like boundary momentum $p_{i}$. For the time-like case, ODEs were solved by imposing the infalling boundary condition on the metric fluctuation, to find its $r$-dependence. For the other two cases, we imposed regularity boundary condition inside the bulk. The results show that the light-like modes do not contribute to the calculation of the two-point functions. On the other hand, the time-like and space-like modes contribute in the same way.

It is readily seen from (\ref{exp}) that in this case we are dealing with an asymptotically $AdS$ space-time which according to the AdS/CFT prescription is dual to a QFT with a UV fixed-point. So the QFT on the boundary for which we have computed different two-point functions is a CFT and hence the computed two-point functions are supposed to fulfill its requirements. 

By looking at the two-point function of energy-momentum tensor obtained in (\ref{2pT}) we observe that it has the same structure as equation (100) in \cite{Coriano:2012wp}. The two-point function of PMR is given in (\ref{pf1intfinal}). The first result of (\ref{pf1intfinal}) is the absence of the scalar degree of freedom, which is in agreement with the results of \cite{Deser:2001us}. We can compare the result of two-point functions for  transverse vector and transverse-traceless tensor degrees of freedom with those of a CFT.
To be more precise, these are our final results for the two-point correlators which all agree with the expectations for a CFT
\bea\label{lablab}
&& \langle \tau_{i j} \tau_{k l}\rangle = A_{\tau}~\hat{\Theta} _{ij,kl}\frac{1}{|x|^{2}} \,,\qquad
\langle V_{i}(x)|V_{j}(0)\rangle = A_{V} \frac{x_{i} x_{j}}{|x|^{4}} \,,\nn\\
&&\langle P^{TT}_{i j}| P^{TT}_{k l}\rangle = A_{T}~\hat{\Theta} _{ij,kl} log|x| \,,
\qquad
\langle\tau_{ij} P_{kl} \rangle =0 \,,\\
&&\qquad \quad \hat{\Theta} _{ij,kl} = \hat{\Theta}_{i k} \hat{\Theta}_{j l}+\hat{\Theta}_{i l}\hat{\Theta}_{j k}-\hat{\Theta}_{i j}\hat{\Theta}_{k l}\,,\nn
\eea
where
\bea
A_{\tau} = 4\pi\,,\qquad A_{V} = -4\pi\,,\qquad A_{T} = -4\pi\,.
\eea
Note that the coefficient, ‎$A_{V}$ is negative which leads to having a negative two-point function for  transverse vector part of PMR.‎
Moreover the scaling dimensions of the transverse-traceless tensor part and transverse vector part of PMR are 2 and 1 respectively which both of them violate the bounds on the scaling dimension of unitary CFT operators\cite{Ferrara:1974,Mack:1977,Minwalla:1997ka, Dobrev:1985vh, Dobrev:1985qv, Dobrev:1985qz}. ‎These observations lead to the conclusion that the dual field theory is non-unitary‎. ‎In \cite{Bender:2007wu} it was claimed that the CG is unitary but after that it was criticized in \cite{Smilga:2008pr,Ilhan:2013xe}‎. ‎Our results which are obtained with a completely different method can be a support for this believe that CG is not unitary at least in the linearized level\footnote{For more discussions about this subject see \cite{Lu:2011ks,Ilhan:2014haa,Joung:2014aba}.}‎.

At the linearized level, it has been shown \cite{Maldacena:2011mk} that one can remove these ghost modes (negative norm states) by imposing Neumann boundary condition. Moreover, the action of CG by imposing this boundary condition, is equivalent to the cosmological Einstein-Hilbert action \cite{Maldacena:2011mk}\footnote{To be more precise, this is equivalent to the cosmological Einstein-Hilbert action plus counter terms. Also for Einstein spaces see \cite{Miskovic:2009bm, Miskovic:2014zja}.}. For sure, checking these observations beyond the linearized level will be very interesting. One of the ways to achieve that is checking the form of higher-point functions (specially three-point functions) of the dual field theory. Raw materials necessary for calculating them are provided in \cite{Grumiller:2013mxa} where in the presence of the sources, the non-linear form of one-point functions of energy-momentum tensor and PMR tensor has been  found. 

The three point functions can be obtained by differentiating of the one-point functions a couple of times with respect to the sources and turning off the sources at the end. Apart from that, having the three-point functions in hand gives us this opportunity to know more about the underlying degrees of freedom of the dual field theory. For example it will help us to study more precisely the operator content and the OPE structure of the dual field theory. We will try to address these questions in our future work. 
 
%%%%%%%%%%%%%%%%%%%%%%%%%%%%%%%%%%%%%%

\appendix
\section{Linearization}
We can write the metric as
\be\label{gdec}
ds^2=\frac{1}{4  r^2} d r^2 + \frac{1}{ r} h_{i j} dx^i dx^j\,, \qquad i,j=1\cdots n\,.
\ee
The following relations will be useful for further calculations.
The Christoffel symbols are given by
\bea
\Gamma^{k}_{i j} &=& \frac12 h^{k a} (-\partial_{a} h_{i j} + \partial_{i} h_{a j} +\partial_{j} h_{i a} )\,,\quad
\Gamma^{i}_{j  r} = -\frac{1}{2 r} h^{i}_{j} +\frac12 h^{i a} h'_{j a}\,,\nn\\
\Gamma^{ r}_{i j} &=& 2 h_{i j} -2  r h'_{i j} \,,\quad
\Gamma^{ r}_{ r  r} = -\frac{1}{ r} \,,\quad
\Gamma^{ r}_{ r j} = \Gamma^{i}_{ r  r} = 0\,.
\eea
The Riemann tensors are
\bea
&& \mathcal{R}_{ r i j k} = \frac{1}{2 r}(\nabla_{k}h'_{ij} -\nabla_{j} h'_{i k}) \,,\cr
&& \mathcal{R}_{ r i  r j} = -\frac{1}{2 r} h''_{ij} +\frac{1}{4 r} h^{a b} h'_{i a} h'_{j b} - \frac{1}{4 r^3} h_{i j} \,,\cr
&& \mathcal{R}_{m i j k} = \frac{1}{ r} R_{m i j k} + \frac{1}{ r} (-h'_{k m} h_{i j} - h'_{i j} h_{k m} + h'_{j m} h_{i k} + h'_{i k} h_{j m})
\cr
&& +\frac{1}{ r^2} (h_{i j} h_{k m} - h_{i k} h_{j m} ) + h'_{i j} h'_{k m} - h'_{i k} h'_{j m} \,,
\eea
and Ricci tensors are given by
\bea
 \mathcal{R}_{ r  r} \!\!&=&\!\! -\frac{n}{4 r^2} -\frac14 h'_{a b}h'^{a b}-\frac12 h''_{a b} h^{a b}\,,\qquad
 \mathcal{R}_{ r i} = -\frac12 \nabla_{i}(h'_{a b} h^{a b}) + \frac12 h^{a b} \nabla_{a}(h'_{i b})\,,\nn\\
 \mathcal{R}_{i j} \!\!&=&\!\! R_{i j} - \frac{n}{ r} h_{ij} + (n-2) h'_{i j} + h_{i j} h'_{a b} h^{a b}
- r (2h''_{ij} - 2 h^{a b} h'_{i a} h'_{j b} + h^{a b} h'_{a b} h'_{i j})\,,
\eea
and finally the scalar curvature is 
\be
\mathcal{R}= r R -n - n^2 -3 r^2 h'_{a b}h'^{a b} -4  r^2 h''_{a b} h^{a b} - r^2 h^{i j} h'_{i j} h^{a b} h'_{a b}
+ 2(n-1)  r h^{a b} h'_{a b} .
\ee
In all relations primes denote the derivative with respect to $ r$ and $R, {R}_{ij}, {R}_{ijmn} $ are constructed out of $h_{ij}$. Here $n$ is the dimension of
boundary space-time.
%%%%%%%%%%%%%%%%%%%%%%%%%%%%%%%%%%%%%%%%%%%%%%%%%%%%%%%%%%%%%%%%%%%%%%%%%%%%
\section{Asymptotic analysis}
To find the asymptotic behavior ($ r \rightarrow 0$) of  the boundary metric we expand the three dimensional metric $h_{i j}$ in the CG coordinate as
\begin{equation}\label{exp}
	h_{i j}=h^{(0)}_{i j} + r^\frac{1}{2} h^{(1)}_{i j}+ r h^{(2)}_{i j}+ r^\frac{3}{2}
	h^{(3)}_{i j}+ r^2 h^{(4)}_{i j}+...\,,
\end{equation}
 If we substitute (\ref{exp}) into the different components of linearized Bach equations (\ref{eomrr}), (\ref{eomij}) and (\ref{eomir}), then up to the linear terms, we obtain the following equations (here, $\nabla$ and $\Box$ are purely constructed out of $h^{(0)}_{i j}$).
The $ r r$ component gives us 
	\begin{eqnarray}\label{eomrr1}
	&&  2 \nabla^{i}{\nabla^{j}{h^{(2)}_{i j}}}-\frac{7}{3} \Box{h^{(2)}} - \frac{5}{2} {R^{(2)}}_{i j} {h^{(0)}}^{i j} - \frac{1}{12} \Box{R^{(0)}}\nn \\
	&&+ r^{\frac{1}{2}} ( 3 \nabla^{i}{\nabla^{j}{h^{(3)}_{i j}}}  - \frac{21}{4} R^{(3)}_{i j} {h^{(0)}}^{i j} - 4 \Box{h^{(3)}} + \frac{3}{4} R^{(3)}- \frac{1}{12} \Box{R^{(1)}}) \nn \\
	&&+ r ( 4 \nabla^{i}{\nabla^{j}{h^{(4)}_{i j}}}  - 6 \Box{h^{(4)}}- 9 R^{(4)}_{i j} {h^{(0)}}^{i j}+ 2 R^{(4)}- \frac{1}{12} \Box{R^{(2)}})+\mathcal{O}( r^{\frac{3}{2}})=0 \,,
	\end{eqnarray}
and the $ij$ component turns out to be
	\begin{eqnarray}\label{eomij1}
	&& r\big( 2 \Box{R^{(0)}_{i j}}-24 h^{(4)}_{i j} - 4 \Box{h^{(2)}_{i j}} -\frac{2}{3} \nabla_{i}{\nabla_{j}{R^{(0)}}} - \frac{2}{3} \nabla_{i}{\nabla_{j}{h^{(2)}}} - 4 \nabla_{i}{\nabla^{a}{h^{(2)}_{j a}}} - 4 \nabla_{j}{\nabla^{a}{h^{(2)}_{i a}}} \nn \\
	&&  + 6 \nabla^{a}{\nabla_{i}{h^{(2)}_{j a}}} + 6 \nabla^{a}{\nabla_{j}{h^{(2)}_{i a}}}\!+\! h^{(0)}_{i j} (8 h^{(4)} \!-\! \frac{1}{3} \Box{R^{(0)}} + \frac{4}{3} \Box{h^{(2)}}
	\!-\!\frac{2}{3} \nabla^{a}{\nabla^{b}{h^{(2)}_{ab}}})\big) \!+\! \mathcal{O}( r^{\frac{3}{2}})=0\,,\nn \\
	&&
	\end{eqnarray}
and finally from the $i r$ component we get
	\begin{eqnarray}\label{eomir1}
	&& r^{\frac{1}{2}} ( \frac{1}{4} \Box{\nabla^{a}{h^{(1)}_{i a}}}- \frac{1}{4} \Box{\nabla_{i}{h^{(1)}}} - \frac{1}{6} \nabla_{i}{R^{(1)}} - \frac{1}{2} \nabla_{i}{h^{(3)}} + \frac{3}{2} \nabla^{a}{h^{(3)}_{i a}}) \nn \\
	&&+  r ( 6 \nabla^{a}{h^{(4)}_{i a}}- \frac{1}{2} \Box{\nabla_{i}{h^{(2)}}} - \frac{1}{3} \nabla_{i}{R^{(2)}} + \frac{1}{2} \Box{\nabla^{a}{h^{(2)}_{i a}}} - 2 \nabla_{i}{h^{(4)}} )
	+\mathcal{O}( r^{\frac{3}{2}})=0\,.
	\end{eqnarray}
Note that in writing the above equations, we have implicitly considered the following expansions for Ricci tensor and scalar curvature
	\begin{eqnarray}
	&&R_{i j}=R^{(0)}_{i j} + r^\frac{1}{2} R^{(1)}_{i j}+ r R^{(2)}_{i j}+ r^\frac{3}{2} R^{(3)}_{i j}+ r^2 R^{(4)}_{i j}+...\,, \nn \\
	&&R=R^{(0)} + r^\frac{1}{2} R^{(1)}+ r R^{(2)}+ r^\frac{3}{2} R^{(3)}+ r^2 R^{(4)}+...\,,
	\end{eqnarray}
where $R^{(0)}_{i j}$ and $R^{(0)}$ are purely constructed out of $h^{(0)}_{i j}$ and for example, $R^{(1)}_{i j}$ and $R^{(1)}$ are determined as follows
\begin{eqnarray}
	&&R^{(1)}_{i j}=-\frac{1}{2} \Box{h^{(1)}_{i j}}+\frac{1}{2} \nabla^{a}{\nabla_{i}{h^{(1)}_{j a}}}+\frac{1}{2}\nabla^{a}{\nabla_{j}{h^{(1)}_{i a}}} -\frac{1}{2} \nabla_{i}{\nabla_{j}{h^{(1)}}}, \nn \\
	&&R^{(1)}=-\Box{h^{(1)}}+ \nabla^{a}{\nabla^{b}{h^{(1)}_{a b}}}\,,
	\end{eqnarray}
and so on.
%%%%%%%%%%%%%%%%%%%%%%%%%%%%%%%%%%%%%%%%%%%%%%%%%%%%%%%%%%
\section{Linearization in transverse-traceless gauge}
	In this appendix we will show that by using the reparametrization invariance of the linearized equations of motion and the Weyl symmetry inherited to the linear order, one can write the following simple form for the linearized equations \cite{Lu:2011ks},\cite{Lu:2011zk}
	\begin{eqnarray}
		(\Box+ 2)(\Box+ 4) \textbf{g}_{\mu \nu} =0\,,
	\end{eqnarray}
where $\textbf{g}_{\mu \nu}$ stands for the metric fluctuations around the $AdS_4$ background $\tb{g}^{(0)}_{\mu\nu}$ 
\be\label{4.2}
g_{\mu \nu}=\tb{g}^{(0)}_{\mu \nu}+\textbf{g}_{\mu \nu}\,.
\ee
In \cite{Lu:2011zk} this analysis has been done for the case of critical gravity, we will use the same analysis for conformal gravity. Recalling no-contribution from the Gauss-Bonnet term to the equations of motion in four dimensions, leads us to conclude that in our case, one can write the equation of motion as follow, where it is related to (\ref{EOM}) by the Bianchi identity
\bea\label{4.3}
&&B_{\mu \nu}=\frac{1}{2}(3R^{ \rho \sigma}R_{ \rho \sigma}-R^{2}+\Box R)g_{\mu \nu} +2R R_{\mu \nu}-6R_{\mu  \rho}{R_{\nu}}^{ \rho}\nn\\
&&-2\nabla_{\mu}\nabla_{\nu}R-3\Box R_{\mu \nu}+3\nabla_{ \rho}\nabla_{\mu}R_{\nu}^{ \rho}+3\nabla_{ \rho}\nabla_{\nu}R_{\mu}^{ \rho}
=0  \,.
\eea
The next step is the linearization of this equation around the $AdS_4$ background, for which we have the subsequent relations
\bea\label{4.4}
&&\tb{R}_{\mu \nu  \rho \sigma}=\frac{\Lambda}{3}(\tb{g}^{(0)}_{\mu  \rho}\tb{g}^{(0)}_{\nu \sigma}-\tb{g}^{(0)}_{\mu \sigma}\tb{g}^{(0)}_{\nu  \rho})\,,\qquad \tb{R}_{\mu \nu}=\Lambda \tb{g}^{(0)}_{\mu \nu}\,,\qquad \tb{R}=4\Lambda \,.
\eea
By inserting (\ref{4.2}) into (\ref{4.3}) and using the equation (\ref{4.4}), we get the following results (the index $L$ indicates being of linear order in metric fluctuations)
\bea\label{4.5}
&&B^{L}_{\mu \nu}=(4\Lambda-3) \mathcal{G}^{L}_{\mu \nu}+\Lambda \tb{g}^{(0)}_{\mu \nu}R^{L}-\tb{g}^{(0)}_{\mu \nu}\Box R^{L}+\nabla_{\mu}\nabla_{\nu}R^{L}=0\,,
\eea
where
\bea
&&\mathcal{G}^{L}_{\mu \nu}=R^{L}_{\mu \nu}-\frac{1}{2}\tb{g}^{(0)}_{\mu \nu}R^{L}-\Lambda \textbf{g}_{\mu \nu}\,, \nn\\
&&R^{L}_{\mu \nu}=-\frac{1}{2}\Box \textbf{g}_{\mu \nu}-\frac{1}{2}\nabla_{\mu}\nabla_{\nu}\textbf{g}+\frac{1}{2}\nabla^{\alpha}\nabla_{\mu}\textbf{g}_{\nu \alpha}+\frac{1}{2}\nabla^{\alpha}\nabla_{\nu}\textbf{g}_{\mu \alpha}\,, \nn\\
&&R^{L}=-\Box \textbf{g}+\nabla^{\mu}\nabla^{\nu}\textbf{g}_{\mu \nu}-\Lambda \textbf{g}\,,
\eea
	Substituting these back into (\ref{4.5}), we finally find to first order in perturbation, the following equation of motion for metric fluctuation $\textbf{g}_{\mu \nu}$ 
	\bea\label{4.8}
	&&B^{L}_{\mu \nu}=\Lambda ( - 3 \Box{\textbf{g}_{\mu \nu}} \!-\! \frac{1}{3}\nabla_{\mu}{\nabla_{\nu}{\textbf{g}}} + 2 \nabla_{\mu}{\nabla^{\alpha}{\textbf{g}_{\nu \alpha}}}  + 2 \nabla_{\nu}{\nabla^{\alpha}{\textbf{g}_{\mu \alpha}}} + \frac{5}{6} \tb{g}^{(0)}_{\mu \nu} \Box{\textbf{g}}- \tb{g}^{(0)}_{\mu \nu}\nabla^{\alpha}{\nabla^{\beta}{\textbf{g}_{\alpha \beta}}}) \nn\\
	&&+ \frac{3}{2} \Box^2{\textbf{g}_{\mu \nu}} +\frac{1}{2} \Box{\nabla_{\mu}{\nabla_{\nu}{\textbf{g}}}} - \frac{3}{2} \Box{\nabla_{\mu}{\nabla^{\alpha}{\textbf{g}_{\nu \alpha}}}} - \frac{3}{2} \Box{\nabla_{\nu}{\nabla^{\alpha}{\textbf{g}_{\mu \alpha}}}} + \nabla_{\mu}{\nabla_{\nu}{\nabla^{\alpha}{\nabla^{\beta}{\textbf{g}_{\alpha \beta}}}}} + \frac{4}{3} \Lambda^{2} \textbf{g}_{\mu \nu}\nn\\
	&&-(\frac{1}{2}  \Box^2{\textbf{g}}- \frac{1}{2}\Box{\nabla^{\alpha}{\nabla^{\beta}{\textbf{g}_{\alpha \beta}}}} + \frac{1}{3} \Lambda^{2}  \textbf{g})\tb{g}^{(0)}_{\mu \nu}=0\,.
	\eea
It is lengthy but straightforward to verify that the above equation is invariant under re-parametrization and Weyl transformation
	\be\label{4.10}
	\textbf{g}_{\mu \nu}\rightarrow \hat{\textbf{g}}_{\mu \nu}=\textbf{g}_{\mu \nu}+\nabla_{\mu}\xi_{\nu}+\nabla_{\nu}\xi_{\mu}+\chi\textbf{g}\,\tb{g}^{(0)}_{\mu \nu}\,.
	\ee
	We aim to show that profiting this freedom, it is possible to choose the metric fluctuation $\hat{\textbf{g}}_{\mu \nu}$ such that it satisfies the transverse-traceless gauge conditions as below 
	\bea\label{4.11}
	\nabla^{\mu}\hat{\textbf{g}}_{\mu \nu}=0\,,\qquad \qquad {\hat{\textbf{g}}}=0\,.
	\eea
	To prove it, we notice that relations (\ref{4.10}) and (\ref{4.11}) together mean that
	\bea
	\nabla^{\mu}\textbf{g}_{\mu \nu}+\chi\nabla_{\nu}\textbf{g}+\Box \xi_{\nu}+\nabla_{\nu}\nabla^{\mu}\xi_{\mu}+\Lambda \xi_{\nu}=0, \qquad	
	(1+4\chi)\textbf{g}+2\nabla^{\mu}\xi_{\mu}=0\,.
	\eea
	Combining both these relations and using the known gauge transformation $\nabla^{\mu}\textbf{g}_{\mu \nu}-\nabla_{\nu}\textbf{g}=0$ we obtain the following result
	\bea
	(\Box+\Lambda) \xi_{\nu}=(\chi-\frac{1}{2})\nabla_{\nu}\textbf{g}\,.
	\eea
	Finally we can use the Weyl freedom to fix $\chi=\frac12$. Therefore, $\xi_{\nu}$ should satisfy a second order differential equation as  
	\bea
	(\Box+\Lambda) \xi_{\nu}=0\,,
	\eea
	which can be readily checked to have a solution at least in our four dimensional case i.e. $\Lambda=-3$.

\section*{Acknowledgments}
We would like to thank Mohsen Alishahiha for collaborations in the early stages of this work and for useful comments and discussions. A.N. and B.K. would like to thank Amir Esmaeil Mosaffa and Iva Lovrekovic for useful discussions. 
The work of A.G. is supported by Ferdowsi University of Mashhad under the grant 2/32139 (1393/08/20).

\end{document}